\documentclass[11pt]{article}

\usepackage[utf8]{inputenc}
\usepackage[T1]{fontenc}
\usepackage[margin=1in]{geometry}
\usepackage{amsmath,amssymb}
\usepackage{graphicx}
\usepackage{booktabs}
\usepackage{array}
\usepackage{multirow}
\usepackage{caption}
\usepackage{subcaption}
\usepackage{xcolor}
\usepackage{xurl}
\usepackage{hyperref}
\hypersetup{
  colorlinks=true,
  linkcolor=blue!50!black,
  citecolor=blue!50!black,
  urlcolor=blue!50!black
}
\usepackage{authblk}
\usepackage{enumitem}
\usepackage{siunitx}
\usepackage{times}
\usepackage{placeins}

\title{\vspace{-2em}\bfseries Wolff-Parkinson-White Detection at 471:1 Class Imbalance: A Leakage-Controlled Study of the Data Bottleneck}

\author[]{Nathael Altman}
\affil[]{Coll\`ege Jean-de-Br\'ebeuf, Montr\'eal, Canada}
\date{}

\begin{document}
\maketitle

\begin{abstract}
\noindent
Wolff-Parkinson-White (WPW) syndrome is a congenital cardiac pre-excitation, clinically important and often missed on the resting 12-lead ECG. Detection is hard: the signature is subtle and the condition rare. We pool two public 12-lead corpora, PTB-XL and Chapman-Shaoxing-Ningbo: 66{,}951 recordings, 142 of them WPW, a prevalence of 0.21\% (about 471:1). Under one pre-specified, leakage-controlled protocol, with a held-out fold contacted exactly once, we compare seven representations of the signal, holding the split and the evaluation fixed. Within these corpora and under a modest compute budget, added diversity and capacity do not raise the ceiling: the most orthogonal detector significantly \emph{hurts}, a feature-union model matches a two-member vote, a convolutional network reaches the wavelet detector without exceeding it, and self-supervised pretraining fails a pre-specified gate. A leak-free learning curve, re-selecting features at every size, still rises at the full 115 positives for the strongest deployed detector (paired 90-to-100\% difference $+0.027$, 95\% CI $[0.019, 0.033]$), so it is not shown to have saturated. An error analysis tested against independent evidence finds that the missed cases have a narrower QRS, confirmed by an on-machine measurement outside our pipeline after we show the sign of this effect depends on which delineator measures it; that uncertain labels show no enrichment among the misses; and that some apparent false positives are recordings the corpus itself codes as pre-excited, placing part of the label problem in the negative class. We measure the optimism of non-nested selection at 0.11 to 0.13 average precision. The deployed output is a percentile rank in a frozen reference distribution, not a probability. On the held-out fold, on 14 positives, it reaches an average precision of 0.595 and an ROC area of 0.950. It is a screening pre-filter, not a diagnostic tool.
\end{abstract}

\vspace{1em}

\section{Introduction}

\subsection{Wolff-Parkinson-White syndrome}

Wolff-Parkinson-White (WPW) syndrome is a congenital cardiac condition in which an accessory atrioventricular conduction pathway allows electrical activation to bypass the normal atrioventricular node and pre-excite the ventricle. On the surface 12-lead electrocardiogram this produces a characteristic triad: a short PR interval (typically below 120 milliseconds), a delta wave (a slow, slurred upstroke at the onset of the QRS complex reflecting early ventricular activation through the accessory pathway), and a consequently widened QRS complex, often with secondary repolarization (ST-T) changes. The clinical importance of recognizing WPW is that the accessory pathway can support re-entrant tachyarrhythmias and, in a subset of patients, predisposes to rapid conduction of atrial fibrillation and a risk of sudden cardiac death; identification on a resting electrocardiogram is therefore consequential, and the pattern is frequently subtle and under-recognized. One terminological point governs the whole paper. What both corpora code, and therefore what we detect, is the WPW electrocardiographic \emph{pattern}: manifest pre-excitation visible on the tracing. The \emph{syndrome} additionally requires documented arrhythmia, which these corpora do not record. We write WPW throughout for readability, but every detection claim in this paper concerns the pattern.

From a signal-processing and machine-learning standpoint, WPW detection has two properties that make it a demanding test case. First, the diagnostic evidence is morphologically subtle: the delta wave is a low-amplitude, short-duration deflection localized to the onset of the QRS, and in cases of minimal or intermittent pre-excitation it may barely widen the QRS at all. Second, WPW is rare. In large unselected electrocardiographic corpora it appears at a prevalence on the order of one in several hundred to one in a thousand recordings, so any realistic detection task is one of extreme class imbalance.

\subsection{The statistical challenge of extreme imbalance}

Extreme class imbalance changes which evaluation metrics are informative. At a positive-to-negative ratio near 471:1, average precision becomes acutely sensitive to the false-positive tail: because the negatives outnumber the positives by three orders of magnitude, a given false-positive rate costs far more precision than it would at balance, so a detector that ranks positives well can still show a modest average precision. This is a sensitivity property, not a ceiling: a perfect ranker attains an average precision of 1.0 at any prevalence. It also means average precision is not comparable across corpora of different prevalence, a point that matters in Section~\ref{sec:batch}. The area under the receiver operating characteristic curve, by contrast, measures ranking quality independently of the base rate. As has been argued for imbalanced settings \cite{saito}, the precision-recall view is the more informative of the two when positives are scarce. We therefore treat average precision as the primary metric for model selection, because it is the more demanding metric under imbalance and is sensitive to the top of the ranked list where a screening tool operates, and we report the area under the receiver operating characteristic curve alongside it as the base-rate-independent summary of discriminative ability. The broader difficulty of learning under skewed class distributions is well documented \cite{hegarcia}.

\subsection{Contribution}

The contribution of this paper is a leakage-controlled study of a rare-class detection problem, in six parts.

\begin{itemize}[leftmargin=1.4em,itemsep=2pt]
  \item \textbf{The binding constraint is data, not model capacity.} We evaluate seven representations of the 12-lead signal under one protocol, holding the split, the evaluation and, where the representation permits it, the admission gate and the learning algorithm fixed (Section~\ref{sec:methods}). Five attempts to raise the ceiling by adding diversity or capacity fail, quantified with paired bootstrap confidence intervals, and a leak-free learning curve, re-running the full feature selection at every training size, shows the strongest deployed detector still improving at the full training set. This is scoped to models trained or pretrained only on these two corpora at modest compute: within that scope the system as deployed is not shown to have saturated, which is weaker than proving data the sole ceiling.
  \item \textbf{A strict anti-leakage discipline.} A fold split that is patient-disjoint wherever patient identity is available, checked directly by a near-duplicate analysis of the positive class (the far more numerous negatives were not exhaustively checked, a lesser risk since the discriminative claims rest on the positives being held out); a held-out test fold contacted exactly once with all model choices frozen in advance; and a label-permutation control whose null collapses to the prevalence.
  \item \textbf{An error and label analysis tested against independent evidence.} In the corpus where an independent measurement exists, the missed cases have a narrower QRS, confirmed by an on-machine measurement outside our pipeline and consistent with minimal pre-excitation; uncertain labels are not enriched among the misses; and part of the label problem lies in the negative class, in recordings the source corpus itself codes as pre-excited. Along the way we show that morphological error analysis is delineator-dependent: a widely used open-source delineator reports the same difference with the wrong sign.
  \item \textbf{A measured bound on selection optimism.} Our feature selection is computed once on the development folds rather than re-nested inside the cross-validation loop, a known source of optimistic bias. On the two models where a fully nested re-run was computationally affordable, that optimism is 0.114 and 0.130 average precision, which calibrates how the out-of-fold numbers throughout this paper should be read.
  \item \textbf{A deliberate deliverable.} We design and justify the deployed output as a percentile rank within a frozen reference distribution rather than a calibrated probability, and characterize the system as a screening pre-filter whose most consequential use is to lower the cost of assembling the larger corpus that would raise its own ceiling.
  \item \textbf{Full reproducibility.} We release the complete decision log, the frozen models, the out-of-fold scores, and the evaluation code, so that the reported numbers can be reproduced from the released artifacts.
\end{itemize}

\FloatBarrier
\section{Related Work}

Automated interpretation of the 12-lead electrocardiogram with machine learning has advanced rapidly, and end-to-end deep networks now approach expert-level performance on several rhythm and morphology tasks, both on large single-institution 12-lead corpora \cite{ribeiro} and on ambulatory single-lead recordings \cite{hannun}. Much of the 12-lead work has been driven by the public release of the PTB-XL corpus \cite{ptbxl} and by community benchmarks built on it. A widely used benchmark on PTB-XL established strong performance across common diagnostic superclasses using one-dimensional convolutional networks, recurrent networks, and wavelet-based front ends, reporting macro-averaged area under the receiver operating characteristic curve with bootstrap confidence intervals \cite{strodthoff}. That benchmark also reports a class-specific figure for WPW itself, of 0.855, which a recent single-lead WPW study adopts as a target baseline \cite{leadi} and which we discuss in Section~\ref{sec:prior}, where the number of positives it rests on turns out to matter a great deal.

Machine-learning work specific to WPW is almost entirely about a different task from ours. It takes the diagnosis as given and localizes the accessory pathway, mapping delta-wave polarity to an ablation site: the classical rule-based algorithm of Arruda and colleagues \cite{arruda}, and its recent deep-learning successors, which reach high accuracy at multi-site localization on independent cohorts \cite{hennecken,senoner}. Localization presupposes that WPW has already been recognized; detection at the corpus base rate, which is our object, is the step before it and is comparatively untreated. This is the gap the present study occupies: not where the accessory pathway is, but whether the pattern is present at all, at a prevalence where the question is statistically hard. Detection has recently been addressed at very large scale, though in a different acquisition setting: a real-world Chinese screening programme scored 3.57 million single-lead recordings from 87{,}836 individuals with a mobile AI-ECG system followed by cardiologist review \cite{huang}. That study asks a deployment and health-economics question rather than a methodological one, and its consumer single-lead signals are not comparable to curated 12-lead archives; we return to it in Sections~\ref{sec:comorbid}, \ref{sec:bottleneck} and \ref{sec:deliverable}, where several of its findings corroborate ours. The single-lead study closest to our task \cite{leadi} examined WPW detection from lead I using transfer learning and wavelet scalograms, motivated by wearable settings; that setting is not directly comparable to the full 12-lead task because it discards most of the spatial information a clinician uses, and, as in the 12-lead benchmarks, external validation across datasets was identified as an open limitation.

The PTB-XL benchmark independently supports one of our design choices: its authors show that splitting a corpus at the level of individual recordings rather than patients systematically overestimates generalization, because a model can exploit training recordings from a patient who also appears in the test set \cite{strodthoff}. We adopt patient-disjoint splitting for exactly that reason, and disclose in Section~\ref{sec:split} the one corpus whose released metadata does not carry patient identifiers.

Three further literatures bear on the specific difficulties of this task. Extreme class imbalance is typically handled with weighted losses, focal reweighting \cite{focalloss}, or augmentation, and evaluation practice increasingly recommends patient-stratified cross-validation and confidence intervals over single point estimates; the danger that model selection performed outside a nested loop inflates the resulting estimate is treated in detail by Cawley and Talbot \cite{cawley}, and we return to it in Section~\ref{sec:feats}. Label noise, pervasive in machine-read ECG corpora, has its own remedies and characteristic failure modes \cite{karimi}, and bears directly on our label analysis in Section~\ref{sec:labelval}. And self-supervised pretraining on unlabeled ECG has been proposed precisely to relieve label scarcity \cite{mehari}, with open 12-lead foundation models now available \cite{ecgfm,hubert}; whether such transfer lifts a rare-class ceiling is one of the questions we test and, within our compute budget, answer in the negative. Cross-institutional validation is repeatedly flagged as missing across this literature, so that reported performance may reflect a single acquisition environment; careful reviews of machine learning in medicine document how easily leakage and inadequate validation inflate apparent performance \cite{roberts}, and broad surveys of deep learning for the ECG map both its opportunities and these recurring validation gaps \cite{hong}. The framing of a rare-class detector as a triage pre-filter rather than a diagnostic instrument follows earlier deep-learning ECG triage systems \cite{vandeleur}. We note as a limitation of the present paper that its engagement with the WPW-specific detection prior art rests on essentially one directly comparable 12-lead reference \cite{leadi}, the one large-scale detection study being single-lead \cite{huang}, and that the rare-event, label-noise, and domain-shift literatures are cited to frame the problem rather than exhaustively surveyed.

\FloatBarrier
\section{Data}

\subsection{Corpora}

We pool two public 12-lead, 10-second electrocardiographic corpora. PTB-XL is a large German corpus \cite{ptbxl}. Chapman-Shaoxing-Ningbo is a large Chinese corpus assembled at Chapman University, Shaoxing People's Hospital, and Ningbo First Hospital \cite{csn_physionet,csn_nature}. The combined dataset contains 66{,}951 recordings, of which 142 carry a WPW label, a prevalence of 0.21\%, corresponding to a class imbalance of approximately 471 to 1. PTB-XL contributes 70 WPW cases among 21{,}799 recordings (a prevalence of 0.321\%) and Chapman-Shaoxing-Ningbo contributes 72 among 45{,}152 (0.160\%). The two corpora therefore differ in WPW prevalence by a factor of two, which is itself worth flagging (it may reflect different labeling practice, different referral populations, or both) and which has direct consequences for how cross-corpus average precision may be compared, treated in Section~\ref{sec:batch}.

The corpora also differ in acquisition hardware, in the analog filtering applied at acquisition, in the underlying patient populations, and in electrode-labeling conventions. These differences produce a strong batch effect: a classifier trained only to distinguish the source hospital, using non-WPW recordings so that the disease cannot be a confound, separates the two corpora with a cross-validated area under the curve of 0.949 on the raw signal. That is a far stronger signal than the WPW signature itself, and quantifying and controlling it is a central methodological concern of this work.

Pooling was a deliberate design decision. A single corpus on its own supplies only about 70 WPW cases, a positive count too small to support reliable model selection: effect-size estimates and the resulting feature choices are unstable at that scale. Adding a second corpus roughly doubles the positive count, at the explicit cost of introducing the batch effect, which we then measure and control explicitly.

\subsection{Split protocol and patient-disjointness}
\label{sec:split}

We use a fixed 10-fold split, stratified so that the scarce positives and the two source corpora are distributed across folds. The split is patient-disjoint in the sense that no patient identifier appears in more than one fold, enforced by a blocking assertion in the pipeline. That guarantee is only as strong as the patient identifiers the corpora provide, so we state precisely what they provide. PTB-XL supplies a genuine patient identifier: its 21{,}799 recordings come from 18{,}869 distinct patients, so 2{,}930 recordings are repeat visits, and the split correctly keeps each patient's recordings together. The Chapman-Shaoxing-Ningbo release supplies no patient identifier at the record level, and this is a documented property of the distributed data rather than a choice; we therefore treat each of its recordings as a distinct patient, and note that record-level indexing is the standard mode of use for this corpus. The Chapman-Shaoxing component is one recording per subject by design; only the Ningbo component could in principle contain repeat recordings, and the release does not let us check.

We test this directly. The only way a missing patient identifier can inflate our results is if the same patient's WPW recording appears in both a training fold and an evaluation fold; because the delta wave is a stable, patient-specific signature, two recordings of the same pre-excited patient would be near-identical in exactly the region the detectors read. We therefore computed the pairwise similarity of all 142 WPW recordings after filtering, using both maximum cross-correlation and cosine similarity on the median beat and on the QRS-onset window. No pair of positives exceeds the near-duplicate threshold, and in particular no held-out (fold-10) WPW recording has a near-twin in the training folds. Near-duplicate leakage within the positive class is thus excluded empirically, independently of the missing identifier; two recordings of the same patient taken years apart need not be near-identical, so this check does not exclude every same-patient pair, only the ones that would matter for a morphology-based detector. This check covers the positives, not the abundant negatives, so a residual effect on the negative side cannot be excluded; but the discriminative claims of this paper rest on the positives being genuinely held out, and for those they are.

Folds 1 through 8 (containing 115 WPW cases) are used for training and model selection and are always reported as pooled out-of-fold performance: for each held-out fold the prediction comes from a model trained on the other seven. Fold 9 (13 WPW) is a validation fold, examined sparingly and never used as the primary basis for a selection decision, because at this positive count its estimates are noisy. Fold 10 (14 WPW among 6{,}713 recordings) is the held-out test set: it is never touched during any development, selection, or tuning, and is contacted exactly once, in a single pre-specified atomic evaluation of the fully frozen system.

\subsection{Label policy and the positive definition}
\label{sec:labelpolicy}

Our positive definition is drawn from each corpus's own coding, and set conservatively. In PTB-XL a recording is positive when the WPW diagnostic statement is present at full likelihood (a likelihood value of 100), which excludes recordings coded WPW at a lower likelihood; this yields 70 positives. In Chapman-Shaoxing-Ningbo a recording is positive when the SNOMED code for Wolff-Parkinson-White (74390002) is present, yielding 72. This is a strict definition; Section~\ref{sec:fp} examines what it excludes and shows the effect on measured performance, which is to make our reported precision conservative.

Beyond the positive definition, our policy is to keep and flag questionable labels rather than to silently remove them: no label is deleted on the basis of a machine-learning intuition, because doing so would risk fitting the evaluation to the model. One further operational definition matters and is easy to conflate. PTB-XL exposes a metadata field recording whether a diagnostic statement was confirmed by a human reader. Throughout this paper, ``non-validated label'' means exactly that field and nothing else: a property of the corpus, established before any model output was examined. Among the PTB-XL WPW positives in the development folds, 44 of 57 (77\%) are non-validated by this criterion, so a non-validated label is the norm rather than the exception in this corpus, a base rate that Section~\ref{sec:labelval} shows is essential to interpreting the error analysis correctly.

\begin{figure}[t]
  \centering
  \includegraphics[width=0.7\linewidth]{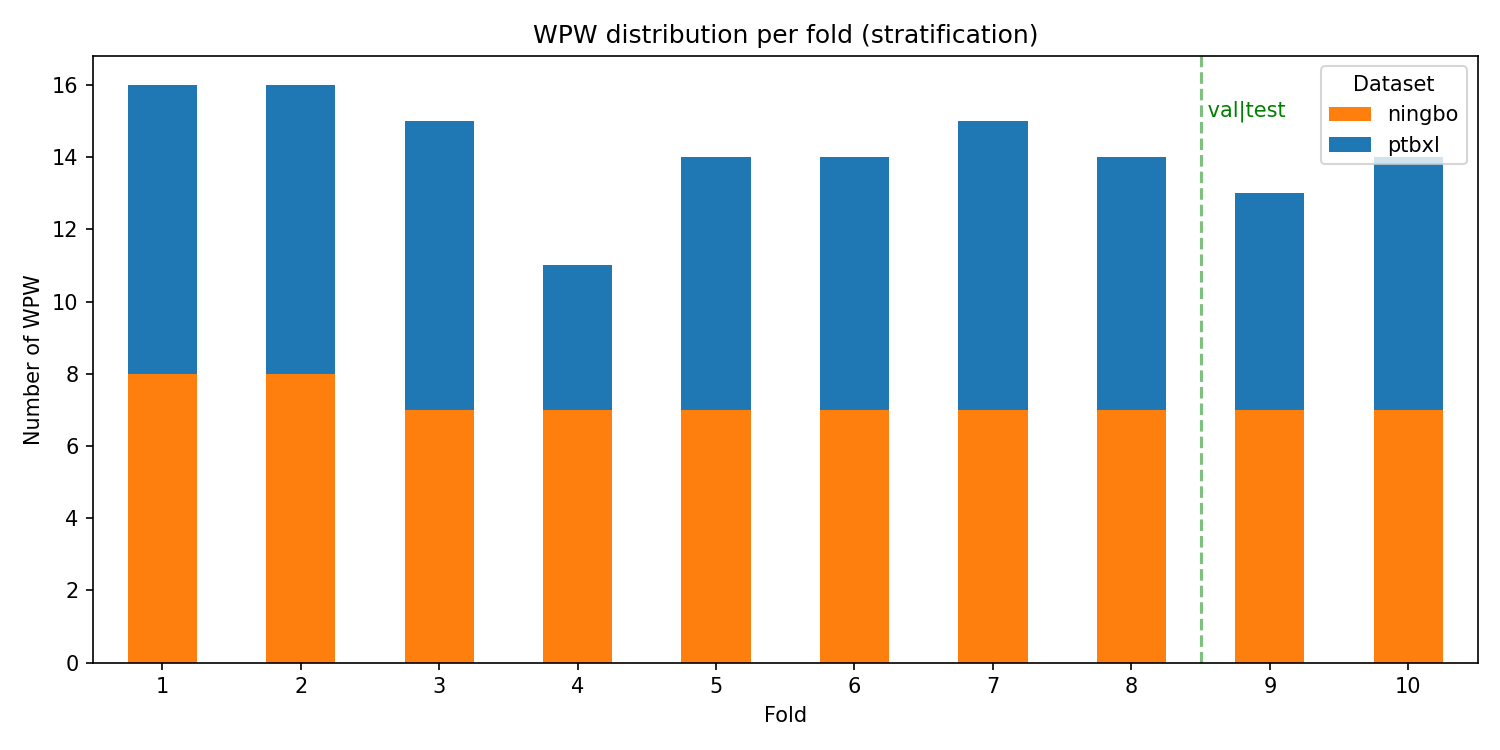}
  \caption{Distribution of WPW cases across the ten folds. Folds 1 through 8 are used for training and out-of-fold selection (115 WPW total), fold 9 for validation (13 WPW), and fold 10 as the single held-out test set (14 WPW). The extreme overall class imbalance (142 WPW among 66{,}951 recordings) is the defining statistical feature of the task.}
  \label{fig:wpwfold}
\end{figure}

\subsection{Cohort composition}
\label{sec:cohort}

Because the positive class is defined by a corpus code rather than by a clinical work-up, it is worth stating who these 142 patients are and what else their tracings carry. The WPW recordings come from a markedly younger population than the archive around them, and the gap appears independently in both corpora: in PTB-XL the WPW median age is 48 years (interquartile range 38 to 60, full range 13 to 87) against 61 years (interquartile range 50 to 72) for the non-WPW recordings, and in Chapman-Shaoxing-Ningbo it is 47 years (interquartile range 28 to 58, range 4 to 83) against 62 years (interquartile range 48 to 72). A congenital accessory pathway is present from birth and is frequently identified before the ages at which acquired cardiac disease accumulates, so this direction is expected. Age therefore confounds every class comparison in this subsection. Sex distribution shows a slight male predominance in the WPW group in both corpora, 54.3\% in PTB-XL and 61.1\% in Chapman-Shaoxing-Ningbo, against 52.1\% and 56.4\% respectively among the non-WPW recordings. PTB-XL censors ages above 89 by storing a fixed sentinel value; 293 non-WPW recordings and no WPW recording are affected, and those are excluded from the PTB-XL age summaries above. Chapman-Shaoxing-Ningbo has 55 non-WPW recordings with no age and 22 with no recorded sex.

Tables~\ref{tab:comorbid-ptbxl} and~\ref{tab:comorbid-csn} give the diagnostic composition of each corpus, WPW recordings against the surrounding archive. Two features are worth naming. First, the WPW recordings carry \emph{fewer} additional abnormalities than the archive, not more: excluding baseline rhythm and normality statements, PTB-XL WPW recordings average 0.84 further statements against 1.59 for non-WPW recordings, and only 1.4\% carry three or more against 25.7\%; in Chapman-Shaoxing-Ningbo the averages are 0.76 against 1.18, with 58.3\% of WPW recordings carrying no further pathological statement at all against 44.1\%. The age gap is sufficient to explain this. The operational consequence is that the typical WPW tracing in these corpora is otherwise clean, which is what makes the minority that is not clean the subgroup of interest in Section~\ref{sec:comorbid}. Second, the statements enriched among WPW recordings are the ones a reader would predict: abnormal QRS in PTB-XL (52.9\% against 15.1\%), supraventricular arrhythmia in PTB-XL (7.1\% against 0.7\%), and supraventricular tachycardia in Chapman-Shaoxing-Ningbo (8.3\% against 1.6\%), the last being the expected accompaniment of an accessory pathway. Conversely the acquired conditions that dominate the archive are depleted or absent among the WPW recordings, including inferior infarction in PTB-XL (2.9\% against 12.3\%) and atrial fibrillation in Chapman-Shaoxing-Ningbo (0 of 72 against 4.0\%).

Three cautions apply to both tables. These are ECG statements attached to a ten-second tracing, not clinical histories: a recording carrying an atrial fibrillation statement had that rhythm on that tracing, and a recording without one may still belong to a patient with paroxysmal disease that the recording did not capture. The two corpora are reported separately and never pooled, because SCP-ECG and SNOMED-CT are different vocabularies and any merge would require an arbitrary mapping. One coded frequency in Table~\ref{tab:comorbid-csn} is worth flagging rather than smoothing over: atrial flutter is coded on 17.9\% of the Chapman-Shaoxing-Ningbo recordings against 4.0\% for atrial fibrillation, a ratio no clinical population would produce. This is a property of that corpus's own coding, not a mapping error on our side, and we report the codes as distributed. And the tables are descriptive; no test is performed on them, the class comparison is confounded by age, and the inferential treatment of comorbidity is in Section~\ref{sec:comorbid}, with its multiplicity correction.

For Chapman-Shaoxing-Ningbo we report the SNOMED-CT preferred term for each code rather than the label shipped in the corpus's own condition-name file, which contains mistranslations and, for two codes, two different names for the same identifier; we resolve names from the PhysioNet/Computing in Cardiology Challenge diagnosis mapping \cite{georgia} and note this because the two sources disagree on 17 of the statements listed here.

\begin{table}[htbp]
\centering
\footnotesize
\caption{PTB-XL diagnostic composition: statements carried by the 70 WPW recordings against the 21{,}729 non-WPW recordings. A statement is listed if it reaches 1\% in either class; statements absent from the WPW group are listed only if they reach 3\% among the non-WPW recordings, since their absence is then informative. Percentages are of the class denominator. Pre-excitation statements are omitted, since they describe the label rather than a concomitant finding. Descriptive only; see Section~\ref{sec:cohort} for the age confound and Section~\ref{sec:comorbid} for the inferential treatment.}
\label{tab:comorbid-ptbxl}
\begin{tabular}{lrrrr}
\toprule
 & \multicolumn{2}{c}{WPW ($n=70$)} & \multicolumn{2}{c}{non-WPW ($n=21{,}729$)} \\
\cmidrule(lr){2-3}\cmidrule(lr){4-5}
Statement & n & \% & n & \% \\
\midrule
Sinus rhythm & 56 & 80.0 & 16692 & 76.82 \\
Abnormal QRS & 37 & 52.9 & 3290 & 15.14 \\
Supraventricular arrhythmia & 5 & 7.1 & 152 & 0.70 \\
Sinus bradycardia & 3 & 4.3 & 634 & 2.92 \\
Anterolateral myocardial infarction & 3 & 4.3 & 285 & 1.31 \\
Inferior myocardial infarction & 2 & 2.9 & 2674 & 12.31 \\
Anteroseptal myocardial infarction & 2 & 2.9 & 2355 & 10.84 \\
Atrial fibrillation & 2 & 2.9 & 1512 & 6.96 \\
Incomplete right bundle branch block & 2 & 2.9 & 1116 & 5.14 \\
Sinus arrhythmia & 2 & 2.9 & 770 & 3.54 \\
Low amplitude T-waves & 1 & 1.4 & 437 & 2.01 \\
Normal ECG & 0 & 0.0 & 9514 & 43.78 \\
Left ventricular hypertrophy & 0 & 0.0 & 2132 & 9.81 \\
Non-diagnostic T abnormalities & 0 & 0.0 & 1825 & 8.40 \\
Left anterior fascicular block & 0 & 0.0 & 1623 & 7.47 \\
Non-specific ischemic & 0 & 0.0 & 1272 & 5.85 \\
Ventricular premature complex & 0 & 0.0 & 1143 & 5.26 \\
Non-specific ST depression & 0 & 0.0 & 1009 & 4.64 \\
Voltage criteria (QRS) for left ventricular hypertrophy & 0 & 0.0 & 875 & 4.03 \\
Sinus tachycardia & 0 & 0.0 & 826 & 3.80 \\
First degree AV block & 0 & 0.0 & 793 & 3.65 \\
Non-specific intraventricular conduction disturbance & 0 & 0.0 & 787 & 3.62 \\
Non-specific ST changes & 0 & 0.0 & 767 & 3.53 \\
Ischemic in anterolateral leads & 0 & 0.0 & 659 & 3.03 \\
\midrule
\multicolumn{5}{p{0.92\linewidth}}{\emph{Statements per record.} All statements: WPW mean 1.64 (2.9\% carry none), non-WPW mean 2.80 (0.0\% carry none). Excluding baseline rhythm and normality statements (\texttt{NORM}, \texttt{SR}): WPW mean 0.84 (37.1\% carry none, 1.4\% carry three or more), non-WPW mean 1.59 (33.4\% carry none, 25.7\% carry three or more).} \\
\bottomrule
\end{tabular}
\end{table}

\begin{table}[htbp]
\centering
\footnotesize
\caption{Chapman-Shaoxing-Ningbo diagnostic composition: statements carried by the 72 WPW recordings against the 45{,}080 non-WPW recordings. A statement is listed if it reaches 1\% in either class; statements absent from the WPW group are listed only if they reach 3\% among the non-WPW recordings. Percentages are of the class denominator. Pre-excitation statements are omitted, since they describe the label rather than a concomitant finding. Statement names are the SNOMED-CT preferred terms resolved from the PhysioNet/Computing in Cardiology Challenge diagnosis mapping \cite{georgia}, not the labels in the corpus's own condition-name file, which disagrees on 17 of the statements below (Section~\ref{sec:cohort}). Descriptive only; see Section~\ref{sec:cohort} for the age confound and Section~\ref{sec:comorbid} for the inferential treatment.}
\label{tab:comorbid-csn}
\begin{tabular}{lrrrr}
\toprule
 & \multicolumn{2}{c}{WPW ($n=72$)} & \multicolumn{2}{c}{non-WPW ($n=45{,}080$)} \\
\cmidrule(lr){2-3}\cmidrule(lr){4-5}
Statement & n & \% & n & \% \\
\midrule
Sinus bradycardia & 32 & 44.4 & 16527 & 36.66 \\
Sinus tachycardia & 13 & 18.1 & 7242 & 16.06 \\
Sinus rhythm & 9 & 12.5 & 8116 & 18.00 \\
Left ventricular high voltage & 9 & 12.5 & 5392 & 11.96 \\
Atrial flutter & 7 & 9.7 & 8053 & 17.86 \\
Sinus arrhythmia & 7 & 9.7 & 2543 & 5.64 \\
Left axis deviation & 6 & 8.3 & 1539 & 3.41 \\
Supraventricular tachycardia & 6 & 8.3 & 718 & 1.59 \\
Complete right bundle branch block & 4 & 5.6 & 1092 & 2.42 \\
Premature ventricular contractions & 3 & 4.2 & 1088 & 2.41 \\
ST changes & 2 & 2.8 & 4230 & 9.38 \\
Premature atrial contraction & 2 & 2.8 & 1310 & 2.91 \\
Q wave abnormal & 2 & 2.8 & 1061 & 2.35 \\
Left bundle branch block & 2 & 2.8 & 238 & 0.53 \\
Paroxysmal ventricular tachycardia & 2 & 2.8 & 107 & 0.24 \\
T wave abnormal & 1 & 1.4 & 7042 & 15.62 \\
Low QRS voltages & 1 & 1.4 & 1042 & 2.31 \\
Right axis deviation & 1 & 1.4 & 852 & 1.89 \\
Nonspecific intraventricular conduction disorder & 1 & 1.4 & 770 & 1.71 \\
Clockwise or counterclockwise vectorcardiographic loop & 1 & 1.4 & 652 & 1.45 \\
Poor R wave progression & 1 & 1.4 & 637 & 1.41 \\
Atrial tachycardia & 1 & 1.4 & 296 & 0.66 \\
Atrial rhythm & 1 & 1.4 & 214 & 0.47 \\
Complete left bundle branch block & 1 & 1.4 & 212 & 0.47 \\
Anterior myocardial infarction & 1 & 1.4 & 56 & 0.12 \\
T wave inversion & 0 & 0.0 & 2877 & 6.38 \\
Atrial fibrillation & 0 & 0.0 & 1780 & 3.95 \\
ST depression & 0 & 0.0 & 1668 & 3.70 \\
\midrule
\multicolumn{5}{p{0.92\linewidth}}{\emph{Statements per record.} All statements: WPW mean 1.61 (0.0\% carry none), non-WPW mean 1.95 (0.0\% carry none). Excluding baseline rhythm and normality statements (sinus rhythm, sinus bradycardia, sinus tachycardia, sinus arrhythmia): WPW mean 0.76 (58.3\% carry none), non-WPW mean 1.18 (44.1\% carry none).} \\
\bottomrule
\end{tabular}
\end{table}

\FloatBarrier
\section{Methods}
\label{sec:methods}

\subsection{Signal filtering}
\label{sec:filter}

All signals are band-pass filtered with a fourth-order Butterworth filter at 0.5 to 40 Hz, applied in a zero-phase (forward-backward) manner. The choice of passband was made empirically, and we state precisely what the evidence for it is and is not, because the comparison predates the detectors it is applied to.

Before any detector had reached its final form, we compared three front ends, no filtering, a 0.5 to 75 Hz band, and the retained 0.5 to 40 Hz band, using out-of-fold discriminative area under the curve (average precision was not computed for this pre-model ablation). The comparison was run on early, simplified precursors of M1, M3, and M7, not on the frozen detectors reported in this paper, and not at all on M4, which did not yet exist. The measured values were, quoted in each case as the retained 0.5 to 40 Hz band first, then the wider 0.5 to 75 Hz band, then no filtering: for the M1 precursor 0.783 against 0.753 and 0.736, a clear preference for the retained band; for the M3 precursor 0.697 against 0.698 and 0.670, which separates the two candidate bands by 0.001 and therefore does not discriminate between them; and for the M7 precursor 0.867 against 0.900 and 0.863, a preference for the wider 0.5 to 75 Hz band.

What the ablation establishes is that filtering helps: all three probes agree on this, and no probe places an unfiltered front end first. What it does not establish is the choice between the two filtered bands. Only the M1 precursor positively designates 0.5 to 40 Hz; on the M3 precursor the two filtered bands differ by 0.001, and the convolutional network's apparent 0.033 preference for the wider band sits well inside its own spread across random seeds, which the frozen record puts at $\pm 0.13$. The retained band is therefore the one band that was positively designated by a probe and never preferred against, and it is the clinical standard for diagnostic electrocardiography; it is not a band shown to be optimal for the models that use it. Three limits remain: the comparison ran on precursors rather than on the frozen detectors, the strongest deployed detector M4 was never tested against an alternative front end, and the ranking between the two filtered bands rests on the physical argument below rather than on a measured difference. Re-running this ablation on the frozen M3 and M4 is a cheap experiment we did not perform.

The physical argument is the following. The delta wave contains spectral energy above 40 Hz, so cutting there discards some of the sharpest high-frequency content of the pre-excitation, but retaining higher frequencies admits substantially more muscle and mains noise, and Chapman-Shaoxing-Ningbo carries roughly 2.6 times the mains power of PTB-XL, so a wider band widens the batch effect as well as the signal. The lower cutoff at 0.5 Hz removes baseline wander while preserving the low-frequency content of the ST segment. A fourth-order rolloff is steep enough to suppress out-of-band noise without the ringing that higher orders introduce. Zero-phase filtering matters specifically because most of our detectors read morphology: a causal filter introduces a frequency-dependent phase shift that distorts the shape and timing of the QRS onset, where the delta wave appears. We tested a causal variant and confirmed the expected onset distortion.

The filter does not solve the batch effect. A hospital classifier trained on non-WPW recordings separates the two corpora with an area under the curve of 0.949 on the unfiltered signal; after the shared 0.5 to 40 Hz band-pass, that separability falls only to approximately 0.90, and no other reasonable band did better. The filter denoises; it does not harmonize the two acquisition environments. This is why our controls rely on explicit cross-dataset evaluation and on a cross-corpus coherence criterion in feature selection, rather than on the belief that a shared filter makes the corpora exchangeable.

\subsection{The seven representations}

Table~\ref{tab:nomenclature} names the seven representations. Each reads a different facet of the 12-lead signal, so that comparison across them isolates the effect of representation as far as the task allows. The fold split and the evaluation routine are identical for all seven. The admission gate is identical for the five feature-based detectors we built; M6 is a single-corpus reference and so cannot receive the cross-corpus coherence criterion, and M7 selects no features at all. The learning algorithm is identical for the six feature-based sets but cannot be for M7, which learns its own representation and is therefore its own learner; that is a limit on the comparison, which we state rather than leave implicit. Two further quantities are set per detector by design and are not held fixed: the Spearman de-duplication threshold (Section~\ref{sec:feats}) and the rule for choosing the final configuration, whose deliberate asymmetry is discussed there.

\begin{table}[t]
  \centering
  \small
  \caption{The seven representations of the 12-lead signal. M6 is a commercial measurement set included as an external reference rather than one of our detectors.}
  \label{tab:nomenclature}
  \begin{tabular}{@{}llp{7.0cm}@{}}
    \toprule
    ID & Descriptive name & Representation \\
    \midrule
    M1 & QRS-onset morphology detector & Beat delineation, retaining onset-morphology and delta-slope descriptors (see text: no classical interval survives selection) \\
    M2 & Global-statistical detector & Per-lead distributional and spectral summaries over the whole 10 seconds, with no beat delineation \\
    M3 & Wavelet-localization detector & Time-frequency wavelet descriptors localized at the QRS onset \\
    M4 & Median-beat morphology detector & Shape of the denoised median beat and of the most pre-excited beat \\
    M5 & Spatial-VCG detector & Vectorcardiographic geometry (Kors and inverse-Dower transforms), delta-axis descriptors \\
    M6 & Commercial baseline & On-machine (Marquette 12SL) measurements; external reference, PTB-XL only \\
    M7 & Convolutional network & Representation learned directly from the raw signal by a one-dimensional residual network \\
    \bottomrule
  \end{tabular}
\end{table}

Detector M1 begins from beat delineation and computes the intervals a clinician reads, notably the PR interval and QRS width, together with delta-slope and wave-morphology descriptors. It is a well-motivated starting point, since these are the very quantities used in the clinical definition of WPW, yet \emph{not one classical interval survives feature selection}. A pool restricted to textbook intervals is empty under our gate (Section~\ref{sec:feats}), because a core interval such as QRS duration separates WPW in opposite directions in the two corpora and so fails the cross-corpus coherence criterion. All 35 features retained by M1 describe the morphology of the QRS onset, the roughly 40 ms window preceding the R peak where the delta wave sits, on leads II, V1, and V5. These intervals are computed by an open-source delineator (NeuroKit2), which Section~\ref{sec:qrsanalysis} shows measures QRS width unreliably on pre-excited beats, even reversing its sign relative to the on-machine Marquette measurement; that the delta wave destabilizes delineation on precisely the pathology the detector targets plausibly contributes to the intervals' failure at the gate, and is consistent with the surviving features being QRS-onset morphology rather than intervals.

Detector M2 summarizes each lead over the full ten seconds using distributional moments, energy bands, and an autocorrelation-based heart-rate estimate, deliberately performing no peak or wave detection. Operating on the whole record rather than on located beats gives it a different failure mode from any delineation-based detector, which is its role in the set.

Detector M3 represents the delta wave as a transient localized at the QRS onset and spread across scales, using a stationary wavelet transform as a shift-invariant core together with wavelet-packet and discrete-wavelet descriptors. Its single strongest feature, and the strongest single feature in the entire study, is a signed wavelet descriptor of QRS-onset polarity with a standardized mean difference of $-2.155$, whose sign flips across leads in a physiologically consistent way and replicates in both corpora separately (PTB-XL $-2.196$, Chapman-Shaoxing-Ningbo $-1.686$). The gate for this detector discards the finest high-frequency detail bands, which is where the batch effect concentrates, without any instruction to do so.

Detector M4 forms a noise-averaged median beat and reads its shape, along with the shape of the most pre-excited beat in the record. That beat is selected by a purely signal-derived criterion, the mean absolute gradient in the window from 80 to 20 ms before the R peak normalized by the peak slope, so the selection never consults the label. It performs its own R-peak detection independently of M1.

Detector M5 reconstructs a vectorcardiogram from the 12 leads using two standard transforms and reads the geometry of the activation loop, which encodes the delta axis a clinician uses to localize the accessory pathway. It is by construction the most orthogonal detector to the others.

Detector M6 uses the on-machine measurements produced by a widely deployed commercial electrocardiographic analysis package, available only for PTB-XL. It is our only external reference, and we report a figure for it in Section~\ref{sec:baselines}, scoped to the corpus where it exists.

Detector M7 is the only representation-learning model in the study: a one-dimensional residual network of roughly 63{,}500 parameters trained from the raw signal. Its architecture, training schedule, and evaluation criteria were fixed before any result was seen, so that its configuration could not be adjusted in response to its own performance. Because the current state of the art in deep ECG analysis often relies on self-supervised pretraining before fine-tuning, we also pretrained this network on the unlabeled development folds and fine-tuned it on the WPW task; that variant is evaluated in Section~\ref{sec:central}. The criterion for accepting the pretrained variant was registered before the run and is stated here so that its failure is checkable: pretraining had to lead the from-scratch network by at least three standard deviations of the seed-to-seed spread, measured on the per-seed mean out-of-fold average precision.
\label{sec:m7gate}

\subsection{Feature selection and the number of features}
\label{sec:feats}

For the feature-based detectors, candidate descriptors number in the hundreds to thousands while the training set contains 115 WPW cases, so unfiltered selection would guarantee false discoveries. Table~\ref{tab:funnel} gives the candidate pool, the number surviving the gate, and the final count for each detector. A feature is admitted only if it passes all of the following simultaneously: a Cohen's standardized mean difference exceeding 0.3 in magnitude; a false-discovery-rate-adjusted significance below 0.05 under Benjamini-Hochberg correction over the full candidate pool \cite{bh}; a bootstrap 95\% confidence interval of the effect size excluding zero; and, for the combined-corpus models, cross-dataset coherence, meaning the same sign and a magnitude above 0.2 within each corpus separately. Surviving features are de-duplicated by removing one of any pair whose Spearman correlation exceeds a threshold, 0.9 for M1 and M2 whose features are relatively independent, and 0.95 for M3, M4, and M5 whose pools are near-collinear by construction. Cross-dataset coherence is the criterion that prevents retaining a feature which only separates WPW in one hospital, and is what stops a model from learning the acquisition environment in place of the disease.

\begin{table}[t]
  \centering
  \small
  \caption{Feature selection funnel per detector, on the pooled development folds. ``Gate'' counts the descriptors passing all four admission criteria simultaneously, before Spearman de-duplication; $K$ is the final count after de-duplication and the joint feature-count/hyperparameter sweep. The gate is stringent: the strongest deployed detector, M4, retains 4.2\% of its candidate pool.}
  \label{tab:funnel}
  \begin{tabular}{@{}lrrrr@{}}
    \toprule
    Detector & Candidate pool & Passing the gate & $K$ final & $K$ / pool \\
    \midrule
    M1 (QRS-onset morphology) & 173   & 73    & 35  & 20.2\% \\
    M2 (Global-statistical)   & 1{,}452 & 756   & 164 & 11.3\% \\
    M3 (Wavelet-localization) & 5{,}004 & 1{,}797 & 500 & 10.0\% \\
    M4 (Median-beat)          & 5{,}184 & 297   & 220 & 4.2\%  \\
    M5 (Spatial-VCG)          & 1{,}570 & 762   & 443 & 28.2\% \\
    \bottomrule
  \end{tabular}
\end{table}

This entire selection procedure is computed once on the pooled development folds (1 through 8), not re-nested within each cross-validation fold. This is precisely the practice Cawley and Talbot identify as a source of optimistic bias \cite{cawley}: the out-of-fold estimates on folds 1 through 8 therefore carry a selection optimism, since each held-out development fold contributed a small share of the univariate and correlation statistics used to choose the feature set. Two structural facts limit how large it can be: selection uses univariate effect sizes and correlation filters rather than the model's own out-of-fold score, and the 115 positives are pooled so any single held-out fold contributes roughly an eighth of the selection statistics.

We can do better than bound it structurally, because for the two feature-union models of Section~\ref{sec:central} a fully nested re-run was computationally affordable, and we performed it. Re-running the entire selection inside each cross-validation fold lowers the union model's average precision from 0.727 to 0.613 (standard deviation 0.004 across repetitions) and the union-plus-intervals model's from 0.740 to 0.610 (standard deviation 0.011). The optimism is therefore on the order of 0.11 to 0.13 average precision for models of this type. Two qualifications matter. These two models draw on the union of all five candidate pools, so they have the most selection freedom of anything in this study and should carry the largest optimism; the single-representation detectors select from smaller pools and are expected to carry less. And this measurement was not repeated for M3 and M4 individually, which would require re-running their extraction inside every fold. The honest reading is that out-of-fold average precisions in this paper should be treated as upper estimates, by an amount that is 0.11 to 0.13 for the most permissive models and unmeasured but presumably smaller for the rest. The selected feature sets were frozen before the single contact with the held-out fold, so the fold-10 result does not carry this optimism at all; only the out-of-fold numbers do.

The gate and the de-duplication decide which features are eligible; they do not decide how many to keep. The surviving features are ranked by the magnitude of their standardized effect size, and a sweep trains a model on the top-ranked $k$ features for $k$ from one to the full de-duplicated pool, recording out-of-fold average precision at each step. The resulting curve rises steeply and then flattens, and the smallest $k$ within a few percent of the curve's maximum is retained as a \emph{provisional} value. That number seeds, but does not determine, the final count: the count is chosen jointly with the learner's tree depth and learning rate in a small grid built around the seed, since the useful number of features depends on the capacity of the model reading them. The rule for picking the winner from that grid differs by detector, and the difference is deliberate. For M1 and M2, a candidate is accepted only if its out-of-fold average precision is statistically tied with the best configuration, judged against the bootstrap confidence interval of that best score, and among tied candidates the smallest feature count is taken. For M3 and M4, the two deployed members, the configuration with the highest out-of-fold average precision wins, with the smaller count breaking any near-tie. The asymmetry is worth stating plainly, because the deployed detectors use the more permissive of the two rules: M1 and M2 were built to characterize what a representation can do, where a parsimony-biased rule guards against reading noise as signal, whereas M3 and M4 were built to be deployed, where the operative question is how well the system can be made to perform and the fold-10 evaluation, frozen in advance, is what checks the answer. The cost of that choice is that M3 and M4 absorb more selection optimism than M1 and M2 in their out-of-fold numbers, which is part of what the nested measurement above quantifies. For M5, an additional cap rejects any candidate whose training average precision runs more than a fixed margin ahead of its out-of-fold value; that cap pulled its final count from 476 to 443 features and its reported out-of-fold average precision from 0.553 to 0.429. We report the lower number. The entire sweep and grid run on folds 1 through 8, with fold 9 visible only as a diagnostic; fold 10 is never involved.

\subsection{Learning algorithm, calibration, and leakage control}

Every feature-based detector (M1 through M6) uses gradient-boosted decision trees \cite{xgboost}; the convolutional network M7 is the single exception and is described below. The decisive property of the tree learner is native handling of missing values: for several detectors, a feature that cannot be computed on a given recording is itself informative, since delineation fails several times more often on a pre-excited beat than on a normal one, and a tree learner that routes missing values natively preserves that signal where a linear model or an imputation step would erase it.

Two alternative learners were tuned as controls under the same protocol, a random forest and a logistic regression, on M1, M2, and M6. On M2, where the comparison was run most carefully with all three learners tuned over their own grids at a fixed feature count, the forest was slightly ahead of the boosted trees out-of-fold (0.306 against 0.266 average precision at that fixed feature count, where the frozen M2 configuration of Table~\ref{tab:hyper} reaches 0.299) and behind on the validation fold (0.303 against 0.407), while the logistic regression plateaued far lower in average precision (0.158 out-of-fold, 0.164 on the validation fold) but attained the highest area under the curve of the three on that fold (0.976; out-of-fold its area under the curve is 0.886, below the forest's 0.931), meaning it ranked reasonably and failed to push WPW cases to the very top, which is the behavior that matters at this base rate. We did not repeat this comparison for M3 and M4, the two deployed detectors, so the choice of learner for the deployed system rests on the missing-value argument and on the M1, M2, and M6 evidence, not on a direct test. Table~\ref{tab:hyper} gives the full configuration of every model.

\begin{table}[t]
  \centering
  \small
  \caption{Configuration of the frozen detectors. All gradient-boosted models share subsample 0.8, colsample-by-tree 0.8, L2 regularization 2.0, minimum child weight 3, and a positive-class scale weight of 464.6 (the inverse base rate). $K$ is the final feature count after selection.}
  \label{tab:hyper}
  \begin{tabular}{@{}lccccc@{}}
    \toprule
    Detector & $K$ & Max depth & Learning rate & Estimators & OOF AP \\
    \midrule
    M1 (QRS-onset morphology) & 35  & 2 & 0.05 & 200 & 0.198 \\
    M2 (Global-statistical)   & 164 & 2 & 0.03 & 300 & 0.299 \\
    M3 (Wavelet-localization) & 500 & 4 & 0.10 & 200 & 0.619 \\
    M4 (Median-beat)          & 220 & 3 & 0.10 & 200 & 0.718 \\
    M5 (Spatial-VCG)          & 443 & 2 & 0.03 & 200 & 0.429 \\
    \bottomrule
  \end{tabular}
\end{table}

The convolutional network M7 is a one-dimensional residual network (convolutional kernel 7, stem kernel 15, batch normalization, dropout), roughly 63{,}500 parameters, trained with Adam at a learning rate of $10^{-3}$, weight decay $10^{-4}$, batch size 64, up to 60 epochs with early stopping at patience 10. We record as a limitation (Section~\ref{sec:limits}) that this network is two orders of magnitude smaller than the architectures used in the PTB-XL benchmark line of work \cite{strodthoff}, and that the reason is a compute constraint rather than a considered modeling choice.

Because the gradient-boosted raw scores are not probabilities and the base rate is extreme, each detector's raw score is passed through a one-input logistic (Platt) fit on the out-of-fold scores. This step does \emph{not} affect average precision or the area under the curve, both of which are invariant to monotone transformations, and the deployed system does not use the calibrated probability at all (Section~\ref{sec:deliverable}). Its function is narrower: it puts the eight per-fold models of a given detector on a common scale before their out-of-fold predictions are pooled into a single curve.

Every model is evaluated through a single shared routine that fixes the operating threshold at the value maximizing the F1 score on the out-of-fold training data, then reports, on held-out data, average precision with a bootstrap confidence interval, area under the curve, the full confusion matrix at that threshold, the multi-seed spread across five random seeds, and the train-minus-out-of-fold gap. One routine for all models means the metrics reported for each are produced by the same code path.

As a direct test for label leakage, we apply a permutation control to each feature-based detector: the labels are shuffled and the out-of-fold pipeline is rebuilt. We used five shuffles per detector, which shows that the null collapses to the prevalence but cannot support a small empirical $p$-value; with five permutations the smallest attainable $p$ is approximately 0.17, which is the strongest statement the run licenses. Section~\ref{sec:perm} reports the collapse, not a significance test. A larger permutation run would license a formal empirical $p$-value, and it is inexpensive; we did not perform it.

\FloatBarrier
\section{Results}
\label{sec:results}

\subsection{Held-out test performance}

\begin{figure}[t]
  \centering
  \includegraphics[width=\linewidth]{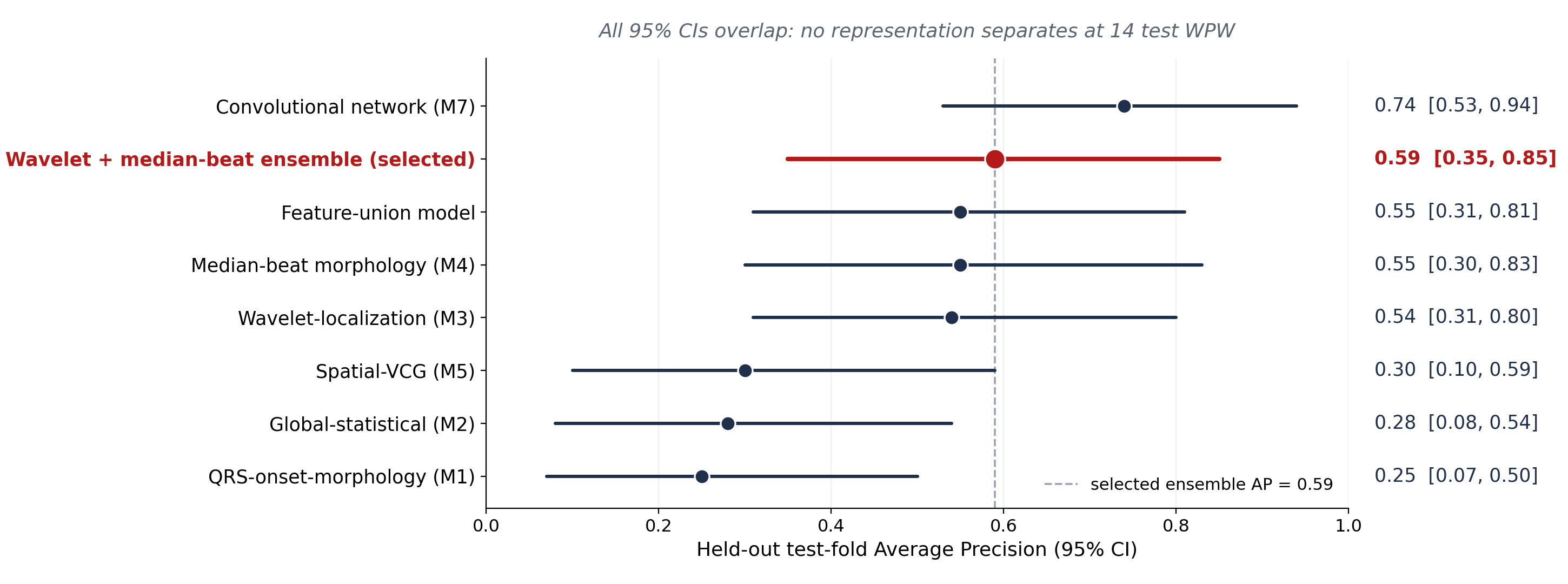}
  \caption{Held-out test-fold average precision with 95\% confidence intervals for every model evaluated on the full 14-positive fold, each thresholded at its own out-of-fold F1-maximizing point (M6 is excluded; see Section~\ref{sec:baselines}). Every confidence interval overlaps every other. At this positive count the ranking between models is not resolvable, which is why no model is selected on this figure.}
  \label{fig:forest}
\end{figure}

\begin{figure}[t]
  \centering
  \includegraphics[width=\linewidth]{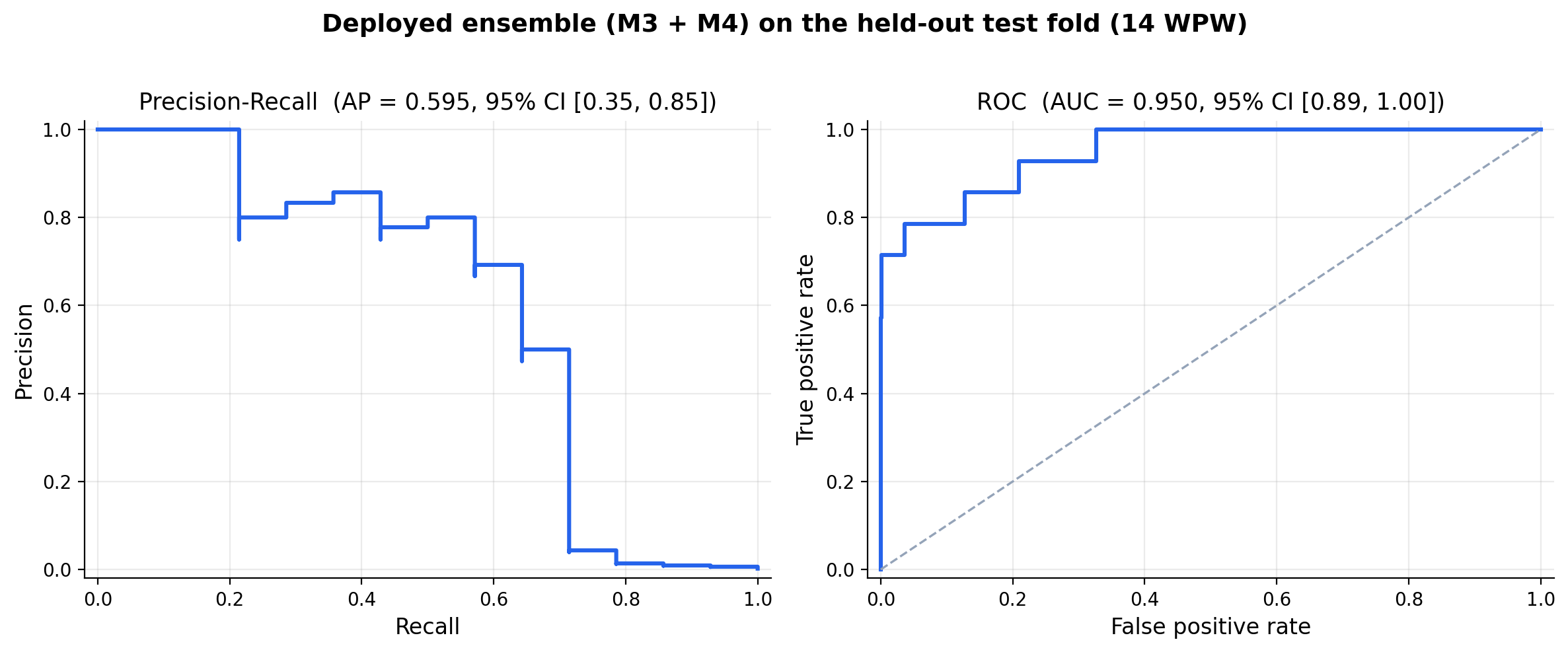}
  \caption{Precision-recall and receiver operating characteristic curves for the deployed ensemble on the held-out test fold (14 WPW). The precision-recall curve shows the steep precision drop at high recall that a 471:1 base rate produces, while the area under the curve of 0.950 shows that ranking quality is high.}
  \label{fig:prroc}
\end{figure}

Figure~\ref{fig:forest} reports the held-out average precision of every model evaluated on the full 14-positive fold, each at its own out-of-fold F1-maximizing threshold; M6 is excluded, for the reason given in Section~\ref{sec:baselines}. The deployed system is a two-member rank-vote of the wavelet-localization and median-beat detectors: each raw score is converted to a percentile against a frozen reference distribution of out-of-fold training scores, the two percentiles are averaged with equal weight, and a single frozen threshold is applied to the average. It attains a held-out average precision of 0.595 (95\% CI [0.346, 0.854]) and an area under the curve of 0.950 (95\% CI [0.89, 1.00]) (Figure~\ref{fig:prroc}). These intervals rest on fourteen positives, and their width bounds what a single held-out contact at this prevalence can establish.

At its frozen operating point it yields 8 true positives, 3 false positives, and 6 false negatives among the 14 held-out WPW cases: a recall of 0.571 and a precision of 0.727. Against the 6{,}699 held-out negatives, 3 false positives give a false-positive rate of 0.045\%, a specificity of 99.96\%. Its out-of-fold average precision on folds 1 through 8 was 0.717, so the held-out estimate sits below the out-of-fold value; at 14 positives, with a confidence interval spanning [0.35, 0.85], this gap is a single undecomposed point and is equally consistent with ordinary sampling variation and with the bounded selection optimism disclosed in Section~\ref{sec:feats}. The data do not separate the two explanations.

We report the threshold-based confusion matrix as an illustration rather than as the deliverable. Section~\ref{sec:batch} shows that absolute score scales are not portable across corpora, and the deployed system therefore operates on percentile ranks, with the operating point a user choice; Section~\ref{sec:deliverable} gives the anchored operating points we actually ship.

\subsection{Why a two-member vote, and why not the convolutional network}
\label{sec:selection}

The deployed system does not beat its own best member on the primary metric, so we state the grounds for deploying it precisely.

On the pooled out-of-fold folds, the equal-weight vote reaches an average precision of 0.717 while M4 alone reaches 0.718: statistically indistinguishable, so the vote is not an average-precision improvement over its strongest member. What it changes is the error profile. At their respective frozen thresholds, the three systems behave as follows on folds 1 through 8 (115 positives):

\begin{table}[h]
  \centering
  \small
  \begin{tabular}{@{}lccccc@{}}
    \toprule
    System & TP & FP & FN & Recall & Precision \\
    \midrule
    M3 alone       & 61 & 21 & 54 & 0.530 & 0.744 \\
    M4 alone       & 87 & 37 & 28 & 0.757 & 0.702 \\
    Deployed vote  & 80 & 25 & 35 & 0.696 & 0.762 \\
    \bottomrule
  \end{tabular}
\end{table}

\noindent Relative to M4 alone, the vote gives up 7 true positives to remove 12 false positives out-of-fold. On the held-out fold the trade is more favorable: M4 alone yields 8 true positives, 10 false positives, and 6 false negatives, against the vote's 8, 3, and 6, so at identical held-out sensitivity the vote produces a third as many false alarms. That is the behavior the composition was chosen for, and it was chosen before fold 10 was opened, by an out-of-fold analysis of where the two detectors make complementary errors.

We deploy the vote for two reasons. The operative one is false-alarm suppression at equal sensitivity, which is the axis a screening aid is judged on. The forward-looking one is robustness to corpus change: a rank-based fusion is the form most likely to survive a shift of acquisition environment, which matters for a future third corpus, though, as Section~\ref{sec:batch} notes, the jointly trained deployed model has already largely harmonized its own scale across the two corpora it was trained on. A reader who weights sensitivity above specificity should prefer M4 alone, and we report both so that choice remains open.

The rank fusion is not the highest-scoring fusion available. Averaging the two members' raw calibrated scores directly reaches an out-of-fold average precision of 0.737, above both the rank vote (0.717) and M4 alone (0.718). We do not deploy it, because averaging raw scores is exactly the operation Section~\ref{sec:batch} shows to be unsafe across acquisition environments: the two corpora place the same detector's scores on scales differing by a factor of 2.7, so a raw-score average is a quantity whose meaning changes with the site. The rank vote gives up roughly 0.02 average precision inside the two corpora it was trained on, in exchange for an output that is defined the same way on a corpus it has never seen. That is the trade, and a reader who cares only about in-distribution average precision should know it was made.

The convolutional network M7 is numerically highest on the held-out fold, and it is deliberately not selected. The decisive evidence is on the out-of-fold folds, where 115 positives give the comparison real power: there M7 reaches an average precision of 0.651 (Table~\ref{tab:five} reports it as $+0.032$ over the wavelet detector M3), below the vote's 0.717 and M4's 0.718. Its numerical lead exists only on the 14-positive held-out fold, where every confidence interval in Figure~\ref{fig:forest} overlaps every other and no ordering is resolvable; promoting a model on that lead would mean selecting on a held-out point estimate the data cannot support. The committee, by contrast, was frozen on the out-of-fold folds on grounds of error complementarity that do not depend on the held-out estimate. We could not run M7's leak-free learning curve either, for the compute reason given in Section~\ref{sec:limits}, so whether it is itself data-limited or capacity-limited is untested. We therefore do not promote it, and we do not read its held-out lead as evidence at this count.

\subsection{The central result: data is the bottleneck}
\label{sec:central}

Five attempts to raise the ceiling by adding representational diversity or model capacity were made, each on the out-of-fold development folds where 115 positives give the comparison some power, and each is reported in Table~\ref{tab:five} as a paired difference in average precision with a bootstrap confidence interval on the difference. Only one is statistically resolved, and it resolves in the wrong direction: adding the most orthogonal detector to the committee significantly \emph{hurts}. Three of the remaining four span zero; for the fifth no paired interval is computable and its point estimates are tied.

\begin{table}[t]
  \centering
  \small
  \caption{Five attempts to raise the ceiling above the two-member vote (out-of-fold average precision 0.717), each a paired bootstrap 95\% confidence interval on the difference (2000 stratified resamples; bounds stable to within 0.003 across three random seeds). The difference is measured against the deployed vote for rows 1--3; for row 4 it is the convolutional network against the wavelet detector M3, its natural engineered baseline, and for row 5 the pretrained network against the same network trained from scratch, as the middle column indicates. These are not five fully independent lines of evidence: rows 2 and 3 are variants of one experiment, as are rows 4 and 5, and row 1 is partly a generic property of committees. The direct test of the thesis is the learning curve (Figure~\ref{fig:learning}).}
  \label{tab:five}
  \begin{tabular}{@{}p{4.4cm}p{4.2cm}p{4.8cm}@{}}
    \toprule
    Attempt & What would confirm it helps & Paired difference in average precision \\
    \midrule
    Add the most orthogonal detector (spatial-VCG) to the committee & A committee gain from the lowest score correlation in the study & $-0.049$, 95\% CI $[-0.080, -0.021]$, excludes zero: the added weak member significantly \emph{hurts} \\
    Feature-union model over all detectors' features & A gain over the vote & $+0.009$, 95\% CI $[-0.026, +0.041]$, spans zero: not separable \\
    Feature-union plus the classical interval features & Clinical intervals add complementary signal & $+0.023$, 95\% CI $[-0.009, +0.055]$, spans zero; these features carry about 4\% of attribution mass \\
    Convolutional network trained from raw signal & A learned representation exceeds engineered ones & $+0.032$ vs the wavelet detector, 95\% CI $[-0.022, +0.084]$, spans zero: reaches the wavelet level, not significantly above \\
    Self-supervised pretraining of that network & Transfer from unlabeled data lifts the scarce-label ceiling & Ensemble point estimates tied (0.625 pretrained vs 0.628 from scratch, under the A/B harness of Section~\ref{sec:m7gate}, which differs from the 0.651 full out-of-fold configuration of row 4). Per seed, pretraining leads by $+0.014$, one standard deviation, against a pre-registered requirement of three: the gate fails \\
    \bottomrule
  \end{tabular}
\end{table}

The feature-union models in rows 2 and 3 are single gradient-boosted models trained on the union of the five detectors' selected features, row 3 additionally including the classical intervals, under the identical protocol; the row-2 model is the one shown as ``Feature-union model'' in Figure~\ref{fig:forest}. These five show that, within the model family we could explore, added diversity and capacity do not raise the ceiling. They do not by themselves prove that data is the binding constraint, and the fact that all the held-out confidence intervals in Figure~\ref{fig:forest} overlap is not evidence for a data-limited regime either: at 14 positives, overlapping intervals are what one would observe under either regime. The direct test is the learning curve, reported in Section~\ref{sec:bottleneck}.

\subsection{Cross-corpus transfer and the batch effect}
\label{sec:batch}

Training on one corpus and testing on the other is the sharpest generalization test the two-corpus structure affords. We report it as a cross-corpus transfer experiment, and deliberately not as ``external validation'': both corpora are training corpora for this study, and in the reporting conventions of clinical prediction models this is internal, not external, validation. (An external check on a third corpus is reported in Section~\ref{sec:external}.) The single-corpus models used here are selected cleanly: their feature gate applies the effect-size, false-discovery, and confidence-interval criteria on the training corpus alone, without the cross-corpus coherence term that the combined models use, so a single-corpus model never consults statistics from the corpus it is about to be tested on. The experiment is therefore uncontaminated by feature selection.

The pattern across detectors is that the area under the curve holds across the crossing (M4, for example, 0.994 within corpus to 0.963 across it) while average precision falls (M4 from 0.693 on PTB-XL to 0.445 tested on Chapman-Shaoxing-Ningbo). Average precision is not comparable across corpora of different prevalence, and this constrains what the drop can be read to mean. Chapman-Shaoxing-Ningbo has half the WPW prevalence of PTB-XL (0.160\% against 0.321\%), and at equal ranking quality that alone roughly halves precision at any recall. Normalizing by the test corpus prevalence, the PTB-XL model's lift, meaning its average precision divided by the prevalence of the corpus it is tested on, \emph{rises} on crossing (216 to 278) while the Chapman-Shaoxing-Ningbo model's lift falls (568 to 195). The drop is therefore partly a prevalence artifact and partly a real degradation, in a direction-dependent way, and it does not on its own establish that score scales fail to transfer.

What does survive, and what actually motivates the rank-based deployment, is a direct measurement of scale non-portability. The raw score distributions of the two corpora are on different scales: M4's 99th-percentile raw score is 0.0336 on PTB-XL and 0.0125 on Chapman-Shaoxing-Ningbo, a factor of 2.7. A threshold read off one corpus's raw scale therefore means something different on the other, whereas the false-positive rate at each corpus's own 99th percentile is identical by construction. This, together with the area under the curve holding across the crossing, is the clean evidence for the design: ranking transfers, absolute scales do not, and so the deployed system fuses and communicates in percentile ranks (Section~\ref{sec:deliverable}). We note for completeness that the deployed model, being trained jointly on both corpora, has largely harmonized its own scale, so its threshold does transfer between them; the scale problem is a property of single-corpus models and of any future third corpus, not a defect visible inside the deployed system itself.

\subsection{No detectable leakage}
\label{sec:perm}

Shuffling the labels and rebuilding the out-of-fold pipeline collapses the average precision to approximately the prevalence for every feature-based detector: 0.198 to 0.003 for M1, 0.299 to 0.002 for M2, 0.619 to 0.002 for M3, 0.718 to 0.002 for M4, and 0.429 to 0.003 for M5, against a base rate of 0.0021. Ranking cannot be reproduced from shuffled labels for any detector. As stated in Section~\ref{sec:methods}, five shuffles per detector establish the collapse but cannot support a $p$-value below approximately 0.17, and we claim only the collapse.

\subsection{External specificity on an unseen corpus}
\label{sec:external}

To check behavior on data from a third acquisition environment never used in training or selection, we ran the frozen deployed system (the M3+M4 rank-vote) over the Georgia 12-lead ECG corpus \cite{georgia}, scoring all 10{,}344 recordings with no ingestion failures. Because that corpus contains only two WPW-labeled recordings, this is a check of specificity and false-alarm behavior, not of sensitivity, and we present it as such. Mapped to the five anchored suspicion levels of Section~\ref{sec:deliverable}, 89.1\% of recordings fall in the lowest level and 10.3\% in the next; only 0.087\% (9 of 10{,}344) reach the two highest levels, and all nine are non-WPW. Among the 1{,}648 recordings carrying a bundle-branch block or intraventricular conduction delay, the conditions most likely to mimic pre-excitation, only two reach the high levels, so the false-alarm rate does not inflate on the population one would most expect to trip it. Of the two WPW-labeled recordings, one is flagged at a moderate level and one is scored in the lowest level. The latter carries a narrow QRS and no visually apparent delta wave, which is an observation about the tracing and not an adjudication of the label: whether it is a mislabel or a case of minimal pre-excitation requires a cardiologist, and no clinical read in this study was performed by one. We therefore count it as a miss. On two positives, external sensitivity is not estimable in either direction, and we make no sensitivity claim from this corpus. The external false-alarm rate, on a corpus the model has never seen, is consistent with its held-out specificity.

\subsection{Non-learned baselines and the commercial reference}
\label{sec:baselines}

Two reference points frame what the learned detectors contribute: the textbook electrocardiographic criteria, and the measurements the acquisition device already produces. They give different answers.

\textbf{The clinical rule.} The textbook electrocardiographic criteria for pre-excitation are a PR interval below 120 ms, a widened QRS above 120 ms, and a delta wave. We can compute the first two from our delineation pipeline; we have no validated automatic delta-wave detector, so we evaluate the two-of-three rule and label it as such. On all 66{,}951 recordings it attains a sensitivity of 0.373 and a specificity of 0.734, flagging 17{,}751 negatives to catch 53 positives, for a precision of 0.003, a lift of only 1.4 over the 0.21\% base rate. Using QRS duration alone as a ranking score gives an area under the curve of 0.551, barely above chance. The rule is unusable at this base rate. Part of this is the rule and part is the delineation it relies on, which Section~\ref{sec:qrsanalysis} shows is unreliable on pre-excited beats; we cannot fully separate the two, but note that a rule built on the same delineation that fails on the target pathology is handicapped from the outset.

\textbf{The commercial measurements (M6).} A detector trained under our protocol on the 782 on-machine Marquette 12SL measurement columns, restricted to the PTB-XL folds where they exist, attains an out-of-fold average precision of 0.583 and an area under the curve of 0.969 on the 57 positives available. The device's own derived measurements therefore carry substantial WPW signal, which is consistent with Section~\ref{sec:qrsanalysis}, where the 12SL QRS duration alone separates detected WPW cleanly. This figure is not directly comparable to the deployed detectors, since it is computed on one corpus, with 57 positives, from a different feature source; but the within-corpus comparison is informative, since on PTB-XL alone the median-beat detector reaches 0.693 (Section~\ref{sec:batch}), modestly above the commercial reference. The value of the learned detectors is thus not that they recover a signal the acquisition device misses, but that they extract somewhat more of it from the raw waveform, and do so on both corpora rather than the single one where the proprietary measurements exist. That the textbook two-of-three rule is useless (precision barely above the base rate) while a model over the same device's measurements is not shows the signal is present in the derived measurements; it is the hand-coded threshold rule, not the measurement set, that fails to exploit it. On the held-out fold, restricted to its 7 PTB-XL positives, M6 reaches an average precision of 0.442; we do not place it in Figure~\ref{fig:forest} because a 7-positive single-corpus estimate is not comparable to the 14-positive figures shown there.

\subsection{Comparison to prior work}
\label{sec:prior}

The closest available comparison point is a WPW discrimination figure of 0.855, cited as a target baseline in a recent single-lead WPW study \cite{leadi} and drawn from the PTB-XL benchmark \cite{strodthoff}. We traced it to its source, and how it is computed determines what can be done with it.

It appears in the benchmark's hierarchical decomposition of class-specific areas under the curve, at the leaf node for WPW, and it is computed on \textbf{eight} WPW cases in that study's test set. It is produced by a single multi-label network trained jointly on all 71 diagnostic statements, with scores propagated up the label hierarchy, rather than by a detector built for WPW; no confidence interval is reported for that node.

The comparison is therefore between two underpowered point estimates. Our held-out area under the curve of 0.950 rests on 14 positives and carries a 95\% confidence interval of [0.89, 1.00]; the reference rests on 8 positives and carries no reported interval, though at that count it would necessarily be very wide. Neither rests on enough positives to support a ranking, so we report the figure for orientation and draw no comparison from the gap between the two numbers. No published 12-lead WPW result, ours included, is estimated on enough positives to rank, which is the scarcity this paper set out to characterize.

\FloatBarrier
\section{Interpretability}
\label{sec:interp}

To check that the detectors read physiology rather than dataset artifacts, we compute Shapley additive explanations \cite{shap} for the five feature-based detectors and Grad-CAM saliency \cite{gradcam} for the convolutional network, on the training folds only.

\begin{figure}[t]
  \centering
  \includegraphics[width=\linewidth]{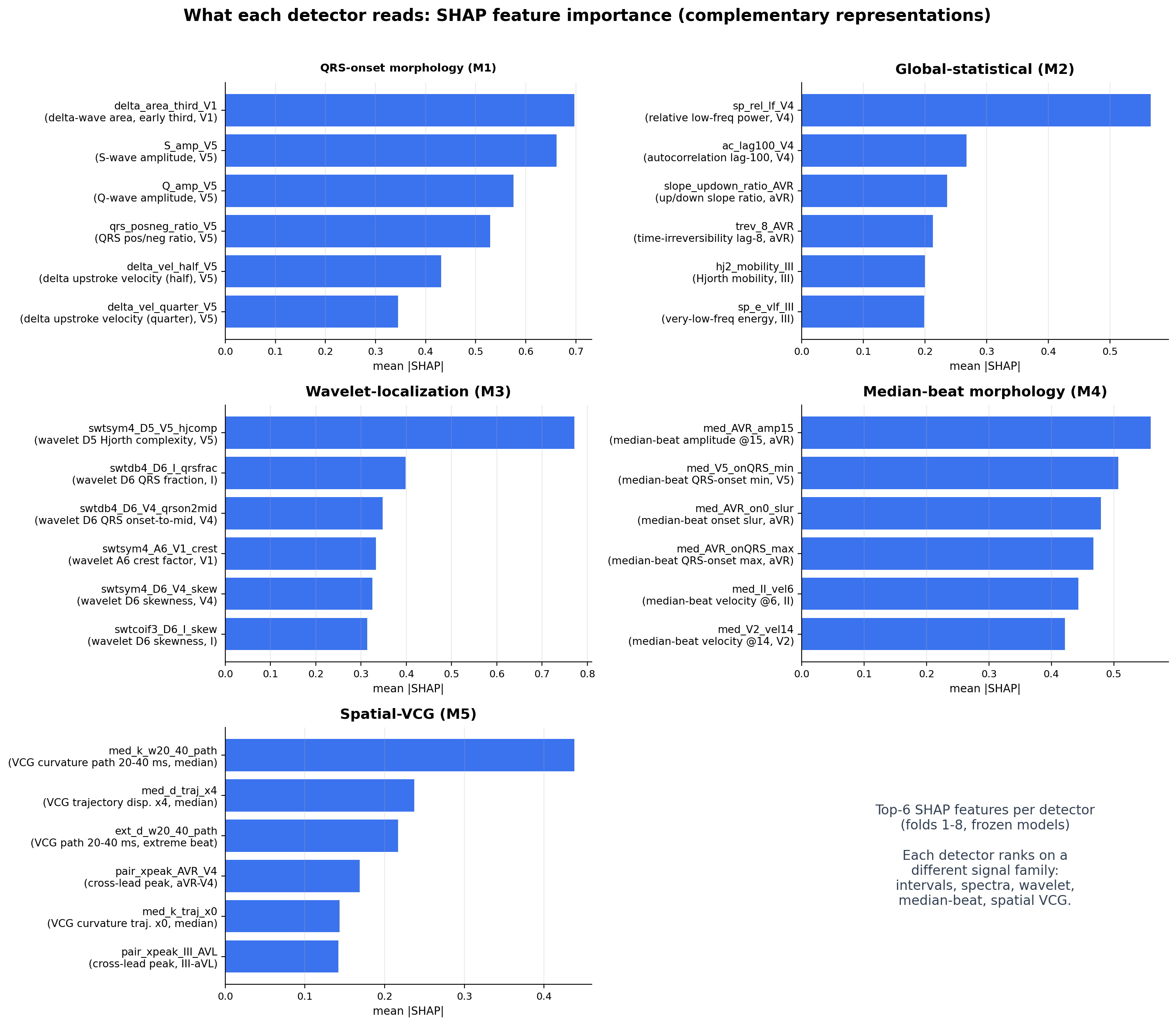}
  \caption{Top features by mean absolute Shapley value for each of the five feature-based detectors (training folds only). Each detector is driven by a distinct facet of the signal and by different leads, which is the mechanistic basis for combining them.}
  \label{fig:shap}
\end{figure}

Each detector's most influential features correspond to a distinct facet of the signal and, notably, to different leads: M1 to a delta-region descriptor in V1, M3 and M4 to morphological descriptors of the QRS in lateral and precordial leads, M5 to activation-loop geometry, and M2 to a spectral descriptor, the least directly interpretable of the five and consistent with its design as a distributional summary (Figure~\ref{fig:shap}).

The single strongest feature by effect size is a signed wavelet descriptor of QRS-onset polarity from M3 (\texttt{swtdb4\_D6\_I\_qrspolsigned}), with a standardized mean difference of about $-2.155$. Its sign flips across leads in a manner consistent with the physiology of ventricular pre-excitation, strongly negative in the high-lateral leads I and aVL and positive in III and aVR, and the signed pattern replicates in each corpus separately. A corpus artifact would not be expected to reproduce across two unrelated acquisition environments with a lead-dependent polarity that matches the expected direction of early ventricular activation.

As a single-feature classifier on the development folds, this descriptor alone reaches an out-of-fold area under the curve of 0.784 and an average precision of 0.096. The area under the curve confirms it carries real discriminative signal on its own; the average precision, far below the full detector's, confirms that it does not by itself constitute a detector at this base rate, so describing M3 as a many-feature model is not misleading. This largest-effect feature is \emph{not} the feature with the largest attribution in Figure~\ref{fig:shap}: the top descriptor by mean absolute Shapley value is a different one, a Hjorth-complexity descriptor on V5 whose own standardized effect is only 0.667. The two need not coincide, and here they do not, because the de-duplication step keeps the highest-effect member of each correlated group while Shapley importance reflects a feature's marginal contribution given all the others; a feature can carry the largest univariate effect and yet rank below a complementary feature once interactions and redundancy are accounted for.

\begin{figure}[t]
  \centering
  \includegraphics[width=0.75\linewidth]{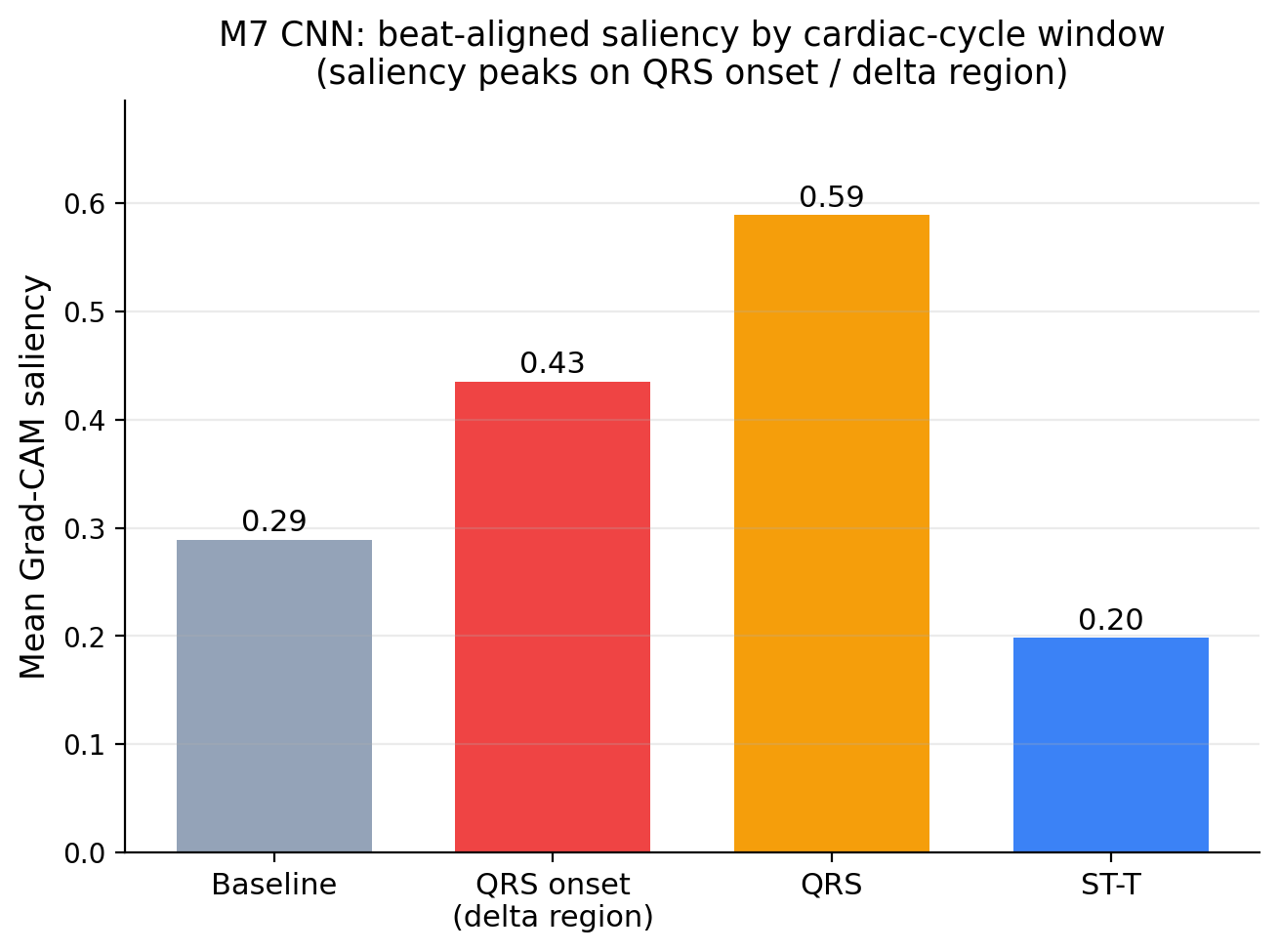}
  \caption{Beat-aligned Grad-CAM saliency of the convolutional network, averaged over true-positive WPW cases and grouped by cardiac-cycle window. Saliency is highest on the QRS and its onset and lowest on the ST-T segment. Because the QRS is the highest-gradient region of any ECG, this pattern is necessary but not sufficient evidence of WPW-specific attention; the control that would make it sufficient was not run.}
  \label{fig:gradcam}
\end{figure}

For M7 we computed Grad-CAM saliency against interpretation criteria registered before the map was produced, since a saliency map read without them is read in whichever direction suits the author. The criteria were: saliency concentrated on the QRS onset would corroborate that the network reads the delta wave, the target of the whole study; saliency on the T wave or ST segment would indicate the network reads secondary repolarization change, a different facet of the WPW picture whose implications for redundancy with M3 and M4 we would then cross-check against the score correlation; and saliency on the record edges or on the isoelectric baseline would indicate an acquisition artifact, to be cross-checked against the corpus-identity confound. The measured saliency is highest on the QRS (0.59) and the QRS-onset delta region (0.43), and lower on the baseline (0.29) and the ST-T segment (0.20) (Figure~\ref{fig:gradcam}), which falls in the first category. This must still be read cautiously. The QRS is the highest-amplitude, highest-gradient region of any electrocardiogram, so saliency peaking there is what almost any ECG classifier would produce for almost any label, and we did not run the control that would make the reading specific, namely the same saliency computed for a network trained on a different diagnosis. What the map does support is the weaker claim that the network is not keying on the baseline or the record edges. We note that the ST-T region receives the least attention of the four windows, even though secondary repolarization change is part of the clinical WPW picture. The network's non-complementarity with M3 and M4 is consistent with all three reading the same region of the beat, but the saliency evidence for that is suggestive, not decisive.

\FloatBarrier
\section{Error Analysis}
\label{sec:error}

Two questions govern this section: which cases the deployed system misses, and whether the explanations available for those misses can be trusted. We test each proposed explanation against an independent source of evidence.

\subsection{Analyzing the right population}
\label{sec:population}

The deployed system is a rank-average vote, not a logical conjunction: a case is flagged when the mean of its two percentile ranks crosses a threshold, so a recording can be flagged with only one member enthusiastic about it, and can be missed although one member flagged it. The errors analyzed here are therefore the committee's own, and not those of either member: 80 true positives, 25 false positives, and 35 false negatives among the 115 development positives, as reported in Section~\ref{sec:selection}.

Of the 35 misses, 14 are in PTB-XL and 21 in Chapman-Shaoxing-Ningbo. Because only PTB-XL carries the label-validation field and the structured comorbidity codes, the decomposition that follows is possible for 14 of the 35 misses, and impossible for the remaining 21. That is a severe limit on the claims this section can make.

\subsection{Missed cases have a narrower QRS, and the measurement is delineator-dependent}
\label{sec:qrsanalysis}

Measured with a delineation-derived QRS-width proxy computed specifically for this analysis (an onset-to-offset width on lead II from a custom R-peak detector, independent of the detectors' own features), the committee's missed cases have a narrower QRS than its detected ones on the PTB-XL positives: a median of 76.5 ms against 118.0 ms (Mann-Whitney $p = 0.0008$; this figure is the PTB-XL restriction of the test, the family member of Table~\ref{tab:holm} being the same descriptor measured over both corpora, at $p = 0.033$). This suggests a straightforward physiological interpretation, that the missed cases are those of minimal or latent pre-excitation, in which conduction through the accessory pathway barely widens the complex.

The reading holds, but establishing it takes a measurement independent of the delineator, because the proxy that suggests it is not itself reliable. PTB-XL is distributed with on-machine Marquette 12SL measurements, including a QRS duration computed by the acquisition device from the same recording, independently of our pipeline. On the same cases the device confirms the narrowing: its QRS duration is a median of 140.0 ms for the detected WPW and 103.0 ms for the missed (Mann-Whitney $p = 2.2\times10^{-6}$). This is the one result in the entire error analysis that survives Holm-Bonferroni correction over the full family of twelve tests reported in Section~\ref{sec:comorbid} (adjusted $p = 2.7\times10^{-5}$), and it is the one measured by an instrument outside our pipeline. Two instruments that disagree on the absolute width thus agree on the direction, and the missed cases genuinely have a narrower QRS; as far as the independent measurement can tell, they are cases of minimal pre-excitation.

\begin{figure}[t]
  \centering
  \includegraphics[width=\linewidth]{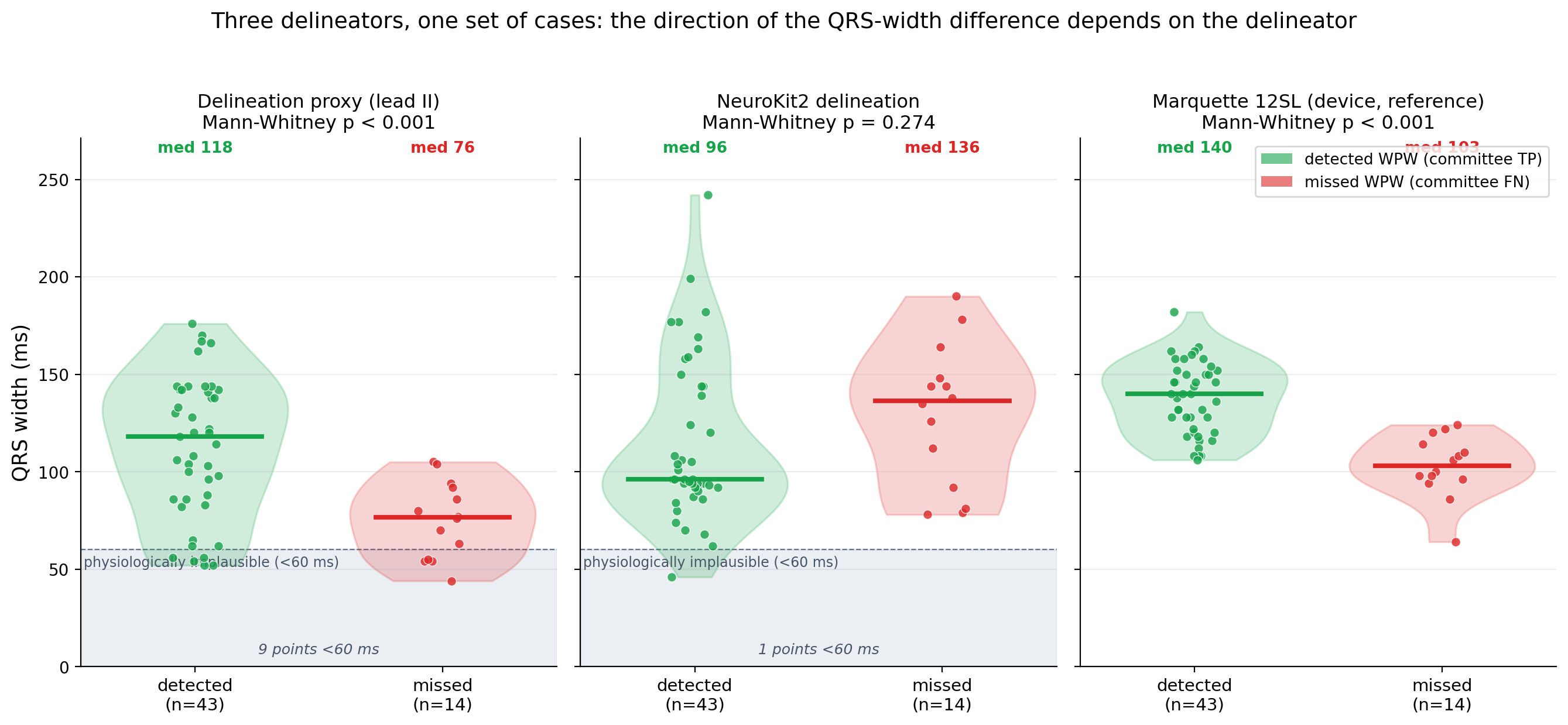}
  \caption{Three measurements of QRS width on the same PTB-XL WPW cases (43 detected, 14 missed by the deployed committee). The delineation proxy built for this analysis (left) and the independent on-machine Marquette 12SL QRS duration (right, the reference) agree that the missed cases have a narrower QRS, consistent with minimal pre-excitation; a widely used open-source delineator (center) reports the opposite. The proxy returns physiologically impossible values below 60 ms on 9 of the 57 cases. The direction of a morphological error-analysis conclusion can depend on which delineator is used, which is why the finding is confirmed against the device rather than asserted from one proxy.}
  \label{fig:qrs}
\end{figure}

What does \emph{not} hold is that any single automatic delineator can be trusted to establish it. Three measurements of QRS width on the same 57 cases disagree (Figure~\ref{fig:qrs}). The device and our custom proxy agree on the direction and correlate at 0.58, but the proxy is quantitatively unreliable: on 9 of the 57 cases it returns a width below 60 ms, not a physiologically possible duration on a resting electrocardiogram, and it understates the device where the two are compared. A widely used open-source delineator, by contrast, \emph{inverts} the relationship: it reports the missed cases as wider (median 136.5 against 96.0 ms), fails to reach significance, and anti-correlates with the device ($-0.22$). Had we run the error analysis with that delineator alone, we would have reported the opposite of the effect the independent instrument measures, with a mechanistically plausible clinical interpretation attached. The same delineation instability appears in Section~\ref{sec:feats}, where no classical interval survives feature selection because delineation fails on precisely the pre-excited beats it is asked to measure. The general form of the problem is not specific to this condition: where the pathology a detector targets also degrades the instrument used to characterize the detector's errors, the resulting error analysis can be internally consistent, mechanistically plausible, and wrong in sign. One corpus here happens to ship an independent measurement that adjudicates. Where none is available, a single-delineator morphological error analysis must be treated as a hypothesis.

\subsection{Comorbidity masking, and a correction for multiple testing}
\label{sec:comorbid}

The strongest surviving signal is comorbidity masking. Among the PTB-XL positives, the committee's missed cases are more likely than its detected ones to carry a QRS-deforming comorbidity, defined in advance as any of bundle-branch block, intraventricular conduction delay, fascicular block, ventricular hypertrophy, or infarction: 5 of 14 misses against 2 of 43 detections (Fisher exact raw $p = 0.0073$, odds ratio 11.4). Unlike the QRS-width result, this one does not depend on a quantity the pathology corrupts; the comorbidity codes are assigned independently of any delineation. The reading is coherent: a condition that itself distorts the ventricular complex can bury the morphological signature of pre-excitation underneath its own.

That said, this section runs twelve hypothesis tests on the same small error population (nine Mann-Whitney descriptor comparisons and three Fisher exact tests), which requires a correction for multiplicity. Table~\ref{tab:holm} lists the full family, since the family size is what determines the correction. Under Holm-Bonferroni over all twelve, exactly one test survives at the 0.05 level, and it is not this one: the QRS duration measured by the acquisition device (Section~\ref{sec:qrsanalysis}) adjusts to $2.7\times10^{-5}$. The comorbidity result adjusts to $p = 0.073$ and does not survive; nor does any other member of the family, whose smallest adjusted value is a beat-count descriptor at 0.0503. Comorbidity masking is the largest effect in the family and the third-smallest raw $p$. We therefore report it as the most promising unconfirmed signal in the error analysis rather than as an established mechanism, and attach no corrected significance to it. At 14 decomposable misses this is the expected state of affairs: the study is not powered to resolve a mechanism, only to rank candidates, and comorbidity masking is the leading candidate. It also tells a clinician something usable even as a hypothesis, namely the population on which the tool is least likely to be reliable. The direction is reproduced outside this study: in the real-world single-lead screening cohort, sensitivity for WPW fell from 48\% in cases without a concomitant cardiac condition to 20\% in cases with one, which the authors attribute to the same mechanism, a complex background obscuring the delta wave \cite{huang}. Both counts are small, so this is convergence and not proof, but it is convergence from an independent population, country, and lead configuration.

\begin{table}[t]
  \centering
  \small
  \caption{The complete family of twelve hypothesis tests run on the committee's error population (80 true positives, 35 false negatives), with raw and Holm-Bonferroni-adjusted $p$-values, ordered by raw $p$. The family is listed in full so that its size, which determines the correction, can be checked rather than taken on trust. Tests marked PTB are restricted to the PTB-XL subset, the only corpus carrying the required metadata. Exactly one test survives correction: the QRS duration measured by the acquisition device. MW denotes Mann-Whitney.}
  \label{tab:holm}
  \begin{tabular}{@{}lrrc@{}}
    \toprule
    Test & Raw $p$ & Holm-adjusted $p$ & Survives \\
    \midrule
    On-machine 12SL QRS duration (MW, PTB) & $2.2\times10^{-6}$ & $2.7\times10^{-5}$ & yes \\
    Beat count (MW) & 0.0046 & 0.0503 & no \\
    QRS-deforming comorbidity, missed vs detected (Fisher, PTB) & 0.0073 & 0.073 & no \\
    Heart rate (MW) & 0.0194 & 0.175 & no \\
    Delta slow-phase duration (MW) & 0.0203 & 0.175 & no \\
    Median QRS width, delineation proxy, both corpora (MW) & 0.0333 & 0.233 & no \\
    Delta-region area (MW) & 0.0514 & 0.308 & no \\
    Bundle-branch block enrichment among false positives (Fisher, PTB) & 0.0526 & 0.308 & no \\
    QRS-width coefficient of variation (MW) & 0.2029 & 0.812 & no \\
    Median R-wave amplitude (MW) & 0.2072 & 0.812 & no \\
    Label validity, validated vs missed (Fisher, PTB) & 0.2706 & 0.812 & no \\
    Normalized delta slope at 40\,ms (MW) & 0.7915 & 0.812 & no \\
    \bottomrule
  \end{tabular}
\end{table}

\subsection{Label validity: a hypothesis that finds no support}
\label{sec:labelval}

A natural hypothesis, given that most of PTB-XL's WPW labels were never confirmed by a human reader, is that the uncertain labels concentrate among the cases the system misses: that the detector is penalized for declining to call recordings that may not be WPW at all. If it held, label validity would be a co-bottleneck alongside label quantity. We tested it directly, and it does not hold.

The test is a Fisher exact on the two-by-two table of validated against non-validated labels by detected against missed, on the PTB-XL positives. Among the 44 non-validated positives, 9 are missed and 35 detected; among the 13 validated positives, 5 are missed and 8 detected. The odds ratio is 0.41 and $p = 0.27$. There is no detectable enrichment, and the point estimate runs the other way: validated cases are missed at a higher rate (5 of 13, 38\%) than non-validated ones (9 of 44, 20\%), though at these counts that is not significant either.

The raw counts are what make the hypothesis look plausible: 9 of the 14 missed PTB-XL cases carry non-validated labels. Against the corpus base rate of 77\% (Section~\ref{sec:labelpolicy}), a missed set that is 64\% non-validated sits \emph{below} the rate at which non-validated labels occur among the positives generally.

Non-validated labels remain a genuine limitation of the corpus, for the reason given in Section~\ref{sec:limits}: a model trained on them partly learns to reproduce an existing automated reader. But they do not explain this system's misses, and label validity is not the co-bottleneck the hypothesis proposes.

\subsection{The label definition, and where it costs precision}
\label{sec:fp}

The false positives are more informative than the false negatives, and they point back at the positive definition itself.

The committee produces 25 out-of-fold false positives, 11 in PTB-XL and 14 in Chapman-Shaoxing-Ningbo. Examining their diagnostic codes:

\begin{itemize}[leftmargin=1.4em,itemsep=1pt]
  \item Chapman-Shaoxing-Ningbo records a separate SNOMED code for ventricular pre-excitation (195060002), distinct from the WPW code (74390002) our positive definition uses. Twelve recordings in the corpus carry it, none of them among our 72 positives. \textbf{Five of the 14 Chapman-Shaoxing-Ningbo false positives carry this ventricular-pre-excitation code.} On the surface electrocardiogram, ventricular pre-excitation and the WPW pattern are closely related; the detector flagged pre-excitation that the corpus documents but that our WPW-specific definition did not count.
  \item \textbf{One of the 11 PTB-XL false positives carries a WPW code at reduced likelihood} (a likelihood value of 50), which our full-likelihood definition excludes. It is a below-threshold WPW label, not a confirmed positive, and we count it only as such.
  \item Among the PTB-XL false positives, QRS-deforming conditions dominate: 7 of the 11 carry a bundle-branch block or intraventricular conduction delay (5) or an inferior infarction (4), with two recordings carrying both. Bundle-branch block is enriched relative to its 20.2\% base rate in PTB-XL negatives (5 of 11, odds ratio 3.29), though at $n = 11$ this is borderline ($p = 0.053$).
\end{itemize}

So 6 of the 25 apparent false positives, 5 documented pre-excitation and 1 below-threshold WPW, are label-definition gaps rather than clear model errors, and most of the genuine remainder are conditions that deform the QRS in ways that mimic pre-excitation, the same mechanism as the misses of Section~\ref{sec:comorbid} acting in the opposite direction. Whether the six are genuinely pre-excited is a labeling determination beyond the scope of this study, and we do not assert it. What we can say precisely is the effect on our reported numbers: under the strict definition these six count against precision, and under an inclusive definition that admits the documented pre-excitation and the below-threshold WPW they would not. Our reported precision and false-positive rate are therefore conservative, and the label problem in this task lies partly in the negative class, in recordings the corpus itself records as pre-excited, rather than only in the positives, where label auditing usually looks.

\subsection{Decomposition and the residual}

The committee's 35 misses can be decomposed only as far as the data permits, which is not far. Of the 14 PTB-XL misses, 9 carry a non-validated label, 3 carry a QRS-deforming comorbidity with a validated label, and 2 are validated, comorbidity-free cases that the system simply missed. The 21 Chapman-Shaoxing-Ningbo misses, 60\% of the total, cannot be decomposed at all, because that corpus carries neither a label-validation field nor comparable structured comorbidity codes.

The residual of 2 is not an estimate of an irreducible physiological floor, and we do not offer it as one. It is the count of PTB-XL misses that neither of our two measurable mechanisms accounts for, drawn from 14 of 35 misses, and one of those mechanisms, label validity, does not discriminate detected from missed cases at all (Section~\ref{sec:labelval}). An unexplained residual of 2 out of 14, in the only decomposable subset, with 21 misses entirely uncharacterized, supports no statement about a physiological floor. What the decomposition does support is narrower and still useful: within the corpus where we can look, a QRS-deforming comorbidity is present in a meaningful share of the misses and is the leading candidate explanation, subject to the multiplicity caveat of Section~\ref{sec:comorbid}, while no other measured quantity separates the misses at all.

\subsection{Reported negative results}
\label{sec:negresults}

Heart rate does not separate the missed from the detected cases on the PTB-XL committee population. Two analysis choices produce an apparently significant result instead, and we record both. Computing the test on a diluted false-negative population, cases missed by at least one member rather than by the committee, produces an effect that the correct population dissolves. And pooling the two corpora produces a difference confounded by source: 60\% of the misses are Chapman-Shaoxing-Ningbo recordings, whose heart-rate distribution differs, so what appears as a heart-rate effect is a corpus effect. That pooled test is the heart-rate member of Table~\ref{tab:holm}, at raw $p = 0.019$, listed there because it was run and not because we read it as an effect. Additional non-discriminant hypotheses include intermittence of pre-excitation, delta-slurring gradient, R-wave amplitude, and per-lead localization.

\FloatBarrier
\section{Discussion}
\label{sec:discussion}

\subsection{Why data, not model, is the bottleneck}
\label{sec:bottleneck}

\begin{figure}[t]
  \centering
  \includegraphics[width=0.8\linewidth]{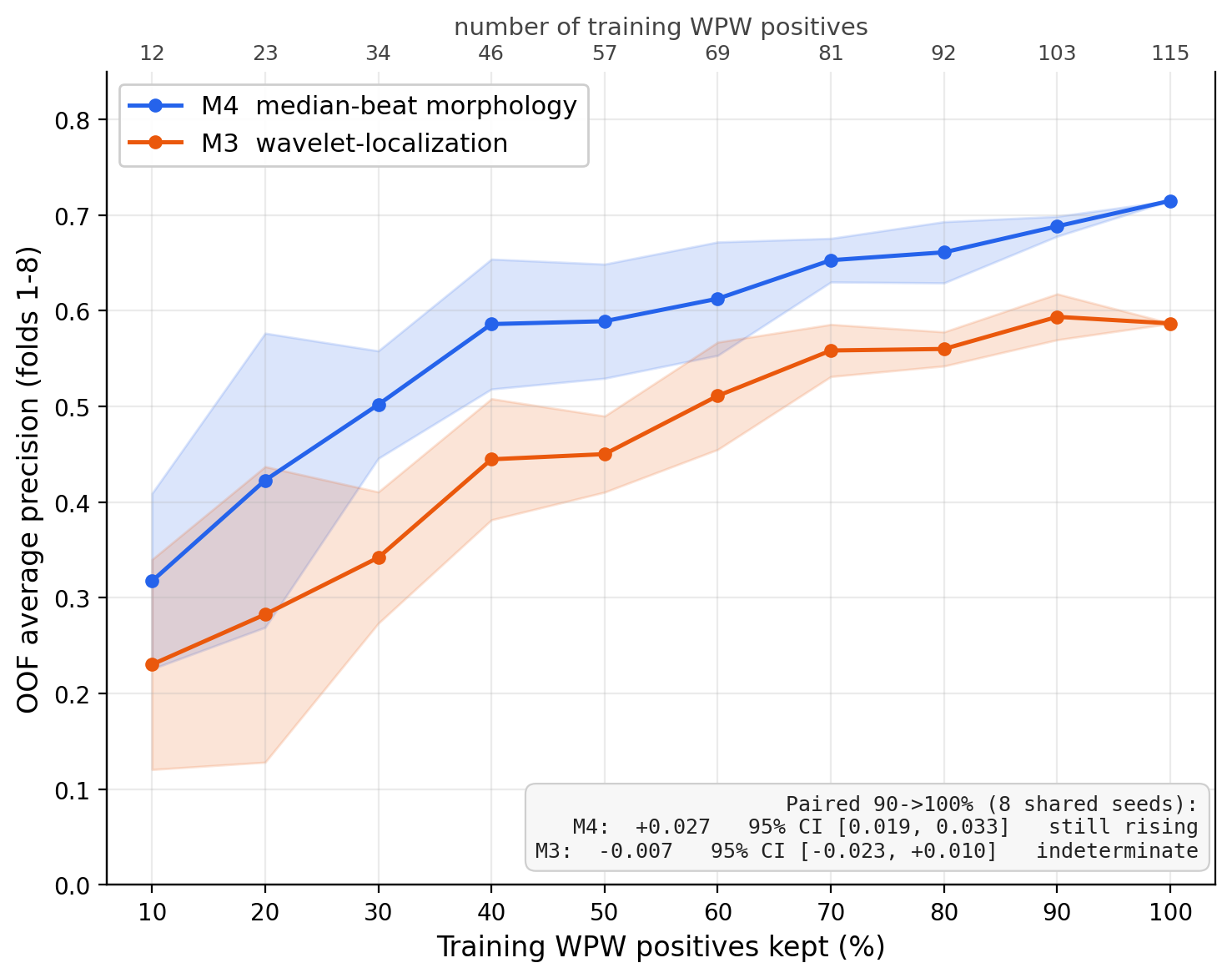}
  \caption{Leak-free learning curves for the two deployed detectors, retrained on increasing random subsets of the training WPW positives (12 to 115 cases) in steps of 10\%, with the negative class held fixed and eight random seeds per point (mean $\pm$ 1 standard deviation, out-of-fold). The full feature selection is re-run from scratch on every subsample, and the subsample is stratified by corpus. Only the positive count is varied, since at 471:1 the negatives are abundant. The median-beat detector (M4) is still rising at the full training set (paired 90-to-100 difference $+0.027$, 95\% CI $[0.019, 0.033]$); the wavelet detector (M3) is inconclusive on its final segment (paired difference $-0.007$, 95\% CI $[-0.023, +0.010]$). The wide seed spread at the smallest fractions, where the gate re-selects on as few as twelve positives, is itself a feature of the low-data regime.}
  \label{fig:learning}
\end{figure}

The five attempts in Table~\ref{tab:five} are convergent but indirect, and, as Section~\ref{sec:central} notes, partly redundant with one another. The direct test is a learning curve on the scarce class. Figure~\ref{fig:learning} retrains the two deployed detectors on increasing random subsets of the training WPW positives, from 10\% to 100\% in steps of 10, holding the negative class fixed and repeating each fraction over eight random seeds. We vary the positives alone, deliberately: at this prevalence the negatives number in the tens of thousands and are not the scarce resource. Two properties of this experiment matter and are treated below: the entire feature selection is re-run from scratch on each positive subsample, so that no low-data point can borrow a feature set chosen with information it did not have; and the positive subsample is stratified by corpus, so that a small fraction cannot collapse onto a single hospital and make the cross-corpus coherence criterion vacuous.

For the median-beat detector (M4, the strongest single member), out-of-fold average precision rises from 0.317 at twelve positives to 0.715 at the full 115 (the learning-curve harness re-selects features at every fraction, which is why the full-data point differs slightly from the frozen 0.718 of Table~\ref{tab:hyper}), and the rise is not finished: the paired difference between the 90\% and 100\% fractions, across the eight shared seeds, is $+0.027$ (standard error 0.004, 95\% CI $[0.019, 0.033]$), which excludes zero. The seed-to-seed standard deviation contracts from 0.092 at the smallest fraction to zero at the full set, where there is nothing left to subsample, the signature of a model still stabilizing as it is given more of the scarce class. This paired test measures the marginal value of the last examples added, not the distance to an asymptote: a positive local slope at the full training set shows the curve has not turned over, but does not bound how much headroom remains. Two properties of the interval should be stated. At the 100\% fraction there is only one possible subsample, so the value is identical across seeds and all of the interval's width comes from the 90\% side; and the eight subsamples at 90\% draw 103 of the same 115 positives, so they overlap heavily and are not independent. The interval is therefore a valid test of whether the mean at 90\% lies below the fixed full-data value, but it is narrower than a fully independent resampling would give.

The wavelet detector (M3) tells a more equivocal story. Its curve rises steeply from 0.230 at twelve positives to about 0.59 by the 70\% fraction, but its final segment is flat: the paired 90-versus-100 difference is $-0.007$ (95\% CI $[-0.023, +0.010]$), which straddles zero and does not establish continued rise. The evidence that this detector is still data-limited at 115 positives is therefore inconclusive.

One inferential step is needed to carry this from the members to the deployed system, and we make it explicit, since we did not compute a learning curve for the rank-average vote itself (the per-fraction per-record member scores needed to reconstruct it were not retained). The vote is a rank average of M3 and M4. Where the two members agree, the vote inherits their common trajectory; where they disagree, the vote's rank on a case tracks whichever member ranks it more extremely. M4 is the stronger and higher-recall member and contributes most of the committee's true positives, so a continued gain in M4 as positives increase raises the committee's ranking on exactly the cases M4 is improving on, unless M3 were already ranking those same cases highly, which its flatter curve makes less likely. The conclusion is thus a bounded one: the deployed system's dominant member is demonstrably still improving with more positives, and there is no member whose saturation would cap the committee, so the committee is not established to be at its ceiling. We do not claim every representation is data-limited, and we do not claim to have measured the ensemble curve directly; we claim that the system as deployed has not been shown to have saturated, and that its strongest component has been shown not to have.

Two notes on the experiment. First, re-running feature selection on each subsample removes the leakage of a fixed feature set but introduces a selection optimism of the opposite sign: on twelve positives the gate picks the features that happen to separate those twelve, which overfits the selection and can inflate a low-data point. The clue is in the seeds. At the 10\% fraction the two protocols disagree in the direction this predicts, the re-selected leak-free curve sitting \emph{above} the fixed-feature curve (0.317 versus 0.225 for M4), and both fall well below the full-data value. Neither protocol is unbiased at low $n$; they are biased in opposite directions; and the qualitative conclusion, that the curve climbs steeply and, for the strongest detector, has not turned over, survives both. Second, we do not extrapolate the curve to a target corpus size: a saturating fit over a range of only 12 to 115 positives is not constrained enough to place an asymptote, and a projected average precision at 1000 positives would dress an unconstrained extrapolation in a false confidence interval. The defensible statement is the one the paired test licenses: at the full training set, the deployed system's strongest member is still improving.

The corollary is usually stated fatalistically: at a prevalence near 471:1, assembling a thousand labeled WPW cases would mean expert review of roughly half a million recordings, so the bottleneck is structural. The arithmetic is right, but it assumes review of an unfiltered archive, and our own specificity result removes that assumption. A detector with a held-out false-positive rate of 0.045\% at a recall near 0.57, or one operating at a recall of 0.93 while flagging 100 recordings per 1000, changes the size of the set a human must read. This has since been measured at scale rather than argued: routing 3.57 million single-lead recordings through an AI pre-filter before cardiologist review cut the number of tracings that must be read to confirm one WPW case from about 1{,}104 to about 18, a roughly sixtyfold reduction in the number of tracings read per case confirmed \cite{huang}. The same study shows why the positives are scarce to begin with: across those 3.57 million recordings, expert review confirmed 55 WPW cases. Scale alone does not yield labeled positives: each must be adjudicated, and rarity is what makes adjudication expensive. The system's value is therefore compounding rather than static. The detector that serves as a screening pre-filter is also the instrument that makes the larger, better-labeled corpus affordable to assemble, and that corpus is the resource the learning curve identifies as still helping the strongest deployed detector at the margin.

\subsection{The deliverable is a rank, not a probability}
\label{sec:deliverable}

How a rare-class detector reports its output is a design decision, and at 471:1 the usual choice is ill-suited. A calibrated probability is a poor deliverable here for two reasons. It is numerically small even for genuinely suspicious tracings, because the base rate pulls every posterior toward zero, so a user shown a single-digit percentage may dismiss a case that warrants review. And it depends on the base rate, which differs between sites, so a threshold calibrated at one institution mislabels risk at another. Section~\ref{sec:batch} shows the deeper version of this: absolute score scales do not transfer across corpora, while rankings do. For the deployed system on its two training corpora this scale problem is already largely internalized, because it was trained jointly on both; the rank output therefore earns its keep chiefly for transfer to a new site and for base-rate portability, the settings in which a probability calibrated at one institution misleads at another. The same conclusion has been reached independently in deployment. In the 3.57-million-recording screening study, the model's absolute probabilities were substantially miscalibrated under ultra-low prevalence, overestimating risk globally, while its ranking behavior held, with confirmed positives concentrating in the top risk stratum; the authors conclude that the model's clinical utility lies in stratification rather than in its probability estimates \cite{huang}, which is the empirical counterpart of the design argument above.

The deployed output is therefore a percentile within a frozen reference distribution of out-of-fold scores, not a probability. A new recording is scored, and its score is expressed as its rank within that fixed reference, a quantity that transfers across acquisition environments even when the raw score does not. Ranks are then mapped to a small number of named suspicion levels anchored on measured operating points rather than round numbers, with the operating level left as the user's choice and a documented per-site recalibration path. Table~\ref{tab:anchors} gives the levels, and gives the quantity a clinician actually needs, which is how many recordings must be reviewed per thousand to achieve a given recall.

\begin{table}[t]
  \centering
  \small
  \caption{The five anchored suspicion levels of the deployed system, measured out-of-fold (folds 1 through 8, 53{,}540 recordings, 115 WPW, prevalence 0.215\%). Each level is the operating point at its lower boundary, so the lowest level admits everything and its precision is the prevalence. ``Flagged per 1000'' is the review burden the level implies. The High level sits one recording away from the frozen operating point of Section~\ref{sec:selection}: its boundary admits 26 false positives for a precision of 0.755, where the frozen threshold admits 25 for 0.762, at the same 80 true positives and the same recall. These points carry the same selection optimism as any out-of-fold estimate (Section~\ref{sec:feats}); only the single operating point near the frozen threshold was confirmed on the held-out fold, so the levels are design targets to be re-measured per site, not validated held-out performance.}
  \label{tab:anchors}
  \begin{tabular}{@{}lccc@{}}
    \toprule
    Level & Precision & Recall & Flagged per 1000 \\
    \midrule
    Very low   & 0.002 & 1.000 & 1000.0 \\
    Low        & 0.020 & 0.930 & 99.9 \\
    Moderate   & 0.300 & 0.809 & 5.8 \\
    High       & 0.755 & 0.696 & 2.0 \\
    Very high  & 0.909 & 0.435 & 1.0 \\
    \bottomrule
  \end{tabular}
\end{table}

The table makes the trade explicit in the terms a screening decision is actually made in. To recover 93\% of the WPW present, roughly one recording in ten must be reviewed. To recover 70\%, two per thousand suffice, and three quarters of what is flagged is real. Which point to occupy is a clinical and economic judgment, not a statistical one, and it is left to the user. Expressing the output as a portable rank, rather than a site-specific probability, is what makes these operating points meaningful across the two corpora, and gives a level definition that carries to a third; the precision and recall attached to each level must be re-measured there.

\subsection{Limitations}
\label{sec:limits}

\textbf{Positives.} 142 WPW cases is a small number, and the held-out confidence intervals are correspondingly wide; at 14 held-out positives no ranking between models is resolvable.

\textbf{Patient-disjointness.} As discussed in Section~\ref{sec:split}, the split is patient-disjoint where a patient identifier exists, and the Chapman-Shaoxing-Ningbo release provides none, so each of its recordings was treated as a distinct patient. The direct near-duplicate check on the 142 positives found no duplication and no held-out twin, which excludes near-duplicate leakage within the positive class; it does not exclude same-patient pairs that are not near-identical, and it does not cover the negatives, so a residual effect on the negative side cannot be formally excluded.

\textbf{Labels.} A fraction of PTB-XL's diagnostic labels are produced by an automated on-machine interpretation and were never confirmed by a human reader, so a model trained on this corpus partly learns to reproduce an existing automated reader \cite{karimi}. Section~\ref{sec:labelval} shows these labels do not explain our misses, but they remain a limit on what the ground truth means, and one corpus carries no validation field at all, which makes 21 of our 35 misses uncharacterizable. Section~\ref{sec:fp} shows the more consequential label problem is in the negative class.

\textbf{Compute, and the capacity claim.} All modeling was performed on a single laptop CPU. The convolutional network we tested has roughly 63{,}500 parameters, two orders of magnitude smaller than the architectures used in the PTB-XL benchmark work we cite \cite{strodthoff}, and the reason is the compute available, not a considered judgment. Our claim that added capacity does not raise the ceiling is therefore established against a small network and against self-supervised pretraining \cite{mehari} on the 53{,}540 unlabeled development-fold recordings, which is a modest corpus for that technique. That pretraining used folds 1 through 8 only, without WPW labels; neither the validation fold nor the held-out fold entered it. The leak-free learning curve, whose per-fraction re-selection is computationally heavy, was run only for the two deployed feature-based detectors; we did not compute it for the convolutional network, whose eighty from-scratch retrainings on CPU were out of reach, so whether the network is itself data-limited or capacity-limited at this scale is untested. We also did not test transfer from a large externally-pretrained ECG foundation model \cite{ecgfm,hubert}, nor supervised transfer from PTB-XL's other diagnostic labels, and either could in principle refute the model-capacity half of our thesis. Every claim in this paper about model sophistication is scoped to models trained or pretrained only on these two corpora.

\textbf{Batch effect.} The batch effect is reduced but not eliminated, and residual confounding cannot be excluded. We do not formally characterize whether the non-WPW populations are clinically comparable across corpora; Section~\ref{sec:fp} finds bundle-branch block enriched among our false positives, so a difference between corpora in the prevalence of QRS-deforming comorbidity among negatives would bear directly on what the detector learns. The cross-corpus coherence gate blocks univariate corpus-identity features but not interaction effects, and the label-permutation control destroys corpus-prevalence signal along with everything else rather than isolating it; a direct check of whether the deployed detector's score on true negatives correlates with corpus identity would close this residual question, and we flag it as a control we did not run.

\textbf{Base rate and the screening setting.} The 0.21\% prevalence is the rate in two hospital archives, not in a screening population, and we do not treat it as a population base rate. A hospital archive over-represents cardiac pathology by construction, because patients are recorded there for a reason, so a general screening population would likely have both a lower WPW prevalence and, more to the point, a lower prevalence of the bundle-branch blocks and infarctions that our false positives concentrate in. The direction this implies is favorable rather than adverse: with fewer QRS-deforming conditions to mimic pre-excitation, the false-alarm profile in a genuine screening population would plausibly be better than the one we report, not worse. We flag this as an untested expectation rather than a result, since it depends on the negative-class composition of a population we did not study. One half of it now has an external figure: the large real-world screening cohort estimates a WPW prevalence near 0.9 per 1{,}000 \cite{huang}, appreciably below the 0.21\% of our pooled hospital archives, which is consistent with archives over-representing the condition.

\textbf{External sensitivity.} The one out-of-distribution check we run (the Georgia corpus, Section~\ref{sec:external}) contains two WPW-labeled recordings, one of which appears to be a mislabel, so it validates specificity and false-alarm behavior only. The defining property of a screening pre-filter is sensitivity, and this study establishes it only in-distribution, on the held-out fold of the two training corpora; external sensitivity is essentially untested, and is the validation we would prioritize before any deployment claim.

\textbf{Scope.} This tool detects manifest pre-excitation on a resting tracing. A concealed accessory pathway, which conducts only retrogradely, produces no delta wave and leaves an entirely normal resting electrocardiogram; it is not a case the tool misses but a case that leaves no trace on the input. Both corpora are overwhelmingly adult: the WPW age range reaches down to 13 years in PTB-XL and 4 in Chapman-Shaoxing-Ningbo (Section~\ref{sec:cohort}), but pediatric recordings are far too few to support any subgroup claim. WPW is disproportionately relevant in the young, and nothing in this study establishes that the tool behaves comparably on pediatric electrocardiograms, which, given the motivation of this work, is the limitation we would most want a reader to carry away.

\textbf{Task framing.} We study direct WPW-versus-rest detection at the corpus prevalence. An alternative strategy trains on a broader conduction-disorder superclass, where the imbalance is far less severe, and reads out the WPW subgroup; it may well detect more WPW in practice. We did not pursue it because it changes the task and removes the imbalance this study is about, so our results say nothing about how it would perform.

\textbf{Method.} De-duplication is applied greedily in order of effect size, so a descriptor redundant with an already-selected feature is dropped even if it would have been more useful in combination. Feature selection was not re-nested inside the cross-validation loop \cite{cawley}. Section~\ref{sec:feats} measures the resulting optimism for the two feature-union models, at 0.114 and 0.130 average precision, but not for M3 and M4 individually, whose nested re-run would require repeating their feature extraction inside every fold; the single-representation detectors draw on smaller candidate pools and are expected to carry less, but that expectation is untested. The same optimism attaches to the development-set comparisons that chose the deployed system, namely the vote composition and the five attempts of Table~\ref{tab:five}, and a nested re-run of the M3-versus-M4-versus-vote comparison is the experiment that would settle it. The single held-out contact is unaffected, since the feature sets and all model choices were frozen before it.

\textbf{Front-end selection.} The passband ablation of Section~\ref{sec:filter} was run on early precursors of M1, M3, and M7, before the deployed detectors reached their final form, and never on M4. It establishes that filtering helps, but it does not establish the specific band on the models that use it: only one precursor positively designates the retained band, and on the others the two filtered bands are separated by less than their own noise. Re-running it on the frozen M3 and M4 is inexpensive and would close the gap.

\textbf{Study design.} This is a retrospective study on curated public corpora. It is not a prospective clinical validation, and every clinical interpretation offered here is a hypothesis requiring clinical confirmation rather than a validated finding.

\subsection{Future work}

The binding constraint is positive examples, so the clear next step is more of them, from additional public corpora and ideally from prospective collection, with the present detector used as a pre-filter to make that collection affordable, which is the compounding use described in Section~\ref{sec:bottleneck}. Clinical confirmation of two specific sets would sharpen the analysis materially: the recordings this study counts as false positives but the source corpus codes as pre-excited (Section~\ref{sec:fp}), and the missed cases carrying a masking comorbidity. Testing supervised transfer from PTB-XL's other diagnostic labels, and transfer from a large externally pretrained ECG model \cite{ecgfm,hubert}, are the two experiments that could refute the model half of our thesis; both are untested here, and neither is assumed unhelpful. Prospective evaluation with per-site recalibration would be required before any clinical use.

\FloatBarrier
\section{Conclusion}

We have presented a leakage-controlled study of Wolff-Parkinson-White detection from the 12-lead electrocardiogram at a class imbalance of approximately 471 to 1. Comparing seven representations of the signal under one protocol, with a held-out test fold contacted exactly once, we find that, scoped to models trained or pretrained only on these two corpora at modest compute, added model diversity and capacity did not raise the ceiling, while a leak-free learning curve is still rising at the full training set for the strongest deployed detector. We therefore conclude that the deployed system has not been shown to have saturated and that its strongest component demonstrably has not, rather than that the quantity of positives has been proven the sole ceiling; the direct evidence is the marginal gain the paired test still detects at the full training set.

The accompanying error and label analysis is tested against independent evidence. In the corpus where an independent measurement exists, the missed cases have a narrower QRS, consistent with minimal pre-excitation; that is the only finding in the analysis to survive multiplicity correction over its full family of twelve tests, and the 21 misses in the other corpus remain uncharacterized. Along the way we find that the sign of the same conclusion depends on which delineator measures it. The hypothesis that uncertain labels concentrate among the misses does not survive comparison against the base rate. And a share of the system's apparent false alarms are recordings the source corpus itself codes as pre-excited, which places part of the label problem in the negative class and makes our reported precision conservative.

The deployed two-member percentile-rank vote reaches a held-out average precision of 0.595 and an area under the curve of 0.950, at a false-positive rate of 0.045\%, on 14 positives. It is a screening pre-filter, not a diagnostic tool, its sensitivity is established only in-distribution, and its most consequential use is the compounding one: it makes affordable the larger, better-labeled corpus that would raise its own ceiling and bring a standalone screening tool within reach. All models, out-of-fold scores, evaluation code, and the complete decision log are released, so that the results here can be reproduced and checked.

\FloatBarrier
\section*{Reproducibility and Code Availability}

The complete implementation is publicly available in the project repository (\href{https://github.com/nathaelaltman/wpw-ecg-detection}{github.com/nathaelaltman/wpw-ecg-detection}), including the frozen model artifacts, the out-of-fold and held-out scores, the per-model evaluation metrics, the interpretability outputs, and a decision log recording every design choice and the alternative it was chosen over. The fold assignment for every recording is released, so the split is reproducible exactly, and the released pipeline reproduces the reported numbers from the frozen artifacts on the pinned dependency stack. Paths are repository-relative and the dependency stack is pinned. Analyses were performed in Python using numpy, pandas, scikit-learn \cite{sklearn}, XGBoost \cite{xgboost}, SHAP \cite{shap}, NeuroKit2 \cite{neurokit}, and PyWavelets \cite{pywavelets}. The held-out test fold was contacted exactly once, and no model choice was made on it.

An interactive demonstration of the frozen deployed system, which scores an uploaded 12-lead recording and returns the suspicion level of Section~\ref{sec:deliverable} together with its feature attributions, is available at \href{https://wpwdetector.com/demo}{wpwdetector.com/demo}. It is a research demonstration of the system described in this paper and is not a medical device, not cleared for clinical use, and not a substitute for interpretation by a qualified clinician.

PTB-XL is publicly available under a Creative Commons Attribution license \cite{ptbxl}, the Chapman-Shaoxing-Ningbo corpus is publicly available through PhysioNet \cite{csn_physionet,csn_nature,physionet}, and the Georgia corpus is publicly available through the PhysioNet/Computing in Cardiology Challenge \cite{georgia}. The commercial on-machine measurements are a proprietary component of the acquisition system and are not redistributed.

\FloatBarrier
\section*{Ethics, Funding, and Competing Interests}

This study uses only publicly released, de-identified corpora whose stated purpose is research use; no new recordings were collected, no participants were recruited, and no institutional ethics approval was required. The work received no funding and was performed on personal equipment. Competing interests: the author was treated for the studied condition at the Montreal Children's Hospital and presents this work publicly alongside a fundraiser for its Foundation; no party to that fundraiser had any role in the design, execution, analysis, or reporting of this study.

\FloatBarrier
\section*{Acknowledgments}

The author thanks the maintainers of the PTB-XL, Chapman-Shaoxing-Ningbo, and Georgia corpora and of the PhysioNet platform for making these data publicly available.


\end{document}